\documentclass{IEEEoj}
\usepackage{cite}
\usepackage{amsmath,amssymb,amsfonts}
\usepackage{algorithmic}
\usepackage{graphicx,color}
\DeclareGraphicsExtensions{.pdf,.png,.jpg}
\usepackage{textcomp}

\usepackage{hyperref}
\usepackage{url}
\usepackage{ragged2e}
\usepackage{multirow}
\let\labelindent\relax
\usepackage{enumitem}
\usepackage{soul}
\usepackage{colortbl} 
\usepackage[table,xcdraw]{xcolor} 
\usepackage{lscape}
\usepackage{lipsum} 
\usepackage{graphicx} 
\usepackage{rotating} 

\usepackage{balance}
\def\BibTeX{{\rm B\kern-.05em{\sc i\kern-.025em b}\kern-.08em
    T\kern-.1667em\lower.7ex\hbox{E}\kern-.125emX}}
\AtBeginDocument{\definecolor{ojcolor}{cmyk}{0.93,0.59,0.15,0.02}}
\def\OJlogo{\vspace{-14pt}\includegraphics[height=28pt]{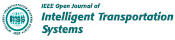}}
\begin{document}
\receiveddate{XX Month, XXXX}
\reviseddate{XX Month, XXXX}
\accepteddate{XX Month, XXXX}
\publisheddate{XX Month, XXXX}
\currentdate{XX Month, XXXX}
\doiinfo{OJITS.2022.1234567}

\title{Interplay between Security, Privacy and Trust in 6G-enabled Intelligent Transportation Systems}

\author{AHMED DANLADI ABDULLAHI\authorrefmark{1} (Student Member, IEEE), ERFAN BAHRAMI\authorrefmark{2}, TOOSKA DARGAHI\authorrefmark{1} (Member, IEEE), MOHAMMED AL-KHALIDI\authorrefmark{1} (Senior Member, IEEE) AND MOHAMMAD HAMMOUDEH\authorrefmark{3} (Senior Member, IEEE)}
\affil{\authorrefmark{1}Department of Computing and Mathematics, Manchester Metropolitan University, Manchester, M15 6BH, United Kingdom}
\affil{\authorrefmark{2}Department of Computer Engineering, Sharif University of Technology, Iran}
\affil{\authorrefmark{3}Department of Information and Computer Science, King Fahd University of Petroleum and Minerals, Saudi Arabia}
\corresp{CORRESPONDING AUTHOR: Ahmed D. Abdullahi (e-mail:ahmed2.abdullahi@stu.mmu.ac.uk).}

\markboth{Interplay between Security, Privacy and Trust in 6G-enabled Intelligent Transportation Systems}{Ahmed \textit{et al.}}

\begin{abstract}
The advancement of 6G technology has the potential to revolutionize the transportation sector and significantly improve how we travel. 6G-enabled Intelligent Transportation Systems (ITS) promise to offer high-speed, low-latency communication and advanced data analytics capabilities, supporting the development of safer, more efficient, and more sustainable transportation solutions. However, various security and privacy challenges were identified in the literature that must be addressed to enable the safe and secure deployment of 6G-ITS and ensure people's trust in using these technologies. This paper reviews the opportunities and challenges of 6G-ITS, particularly focusing on trust, security, and privacy, with special attention to quantum technologies that both enhance security through quantum key distribution and introduce new vulnerabilities. It discusses the potential benefits of 6G technology in the transportation sector, including improved communication, device interoperability support, data analytic capabilities, and increased automation for different components, such as transportation management and communication systems. 
A taxonomy of different attack models in 6G-ITS is proposed, and a comparison of the security threats in 5G-ITS and 6G-ITS is provided, along with potential mitigating solutions. 
This research highlights the urgent need for a comprehensive, multi-layered security framework spanning physical infrastructure protection, network protocol security, data management safeguards, application security measures, and trust management systems to effectively mitigate emerging security and privacy risks and ensure the integrity and resilience of future transportation ecosystems.
\end{abstract}



\maketitle

\section{INTRODUCTION}
\begin{figure}[!htb]
    \centering
     \includegraphics[width=1\linewidth]{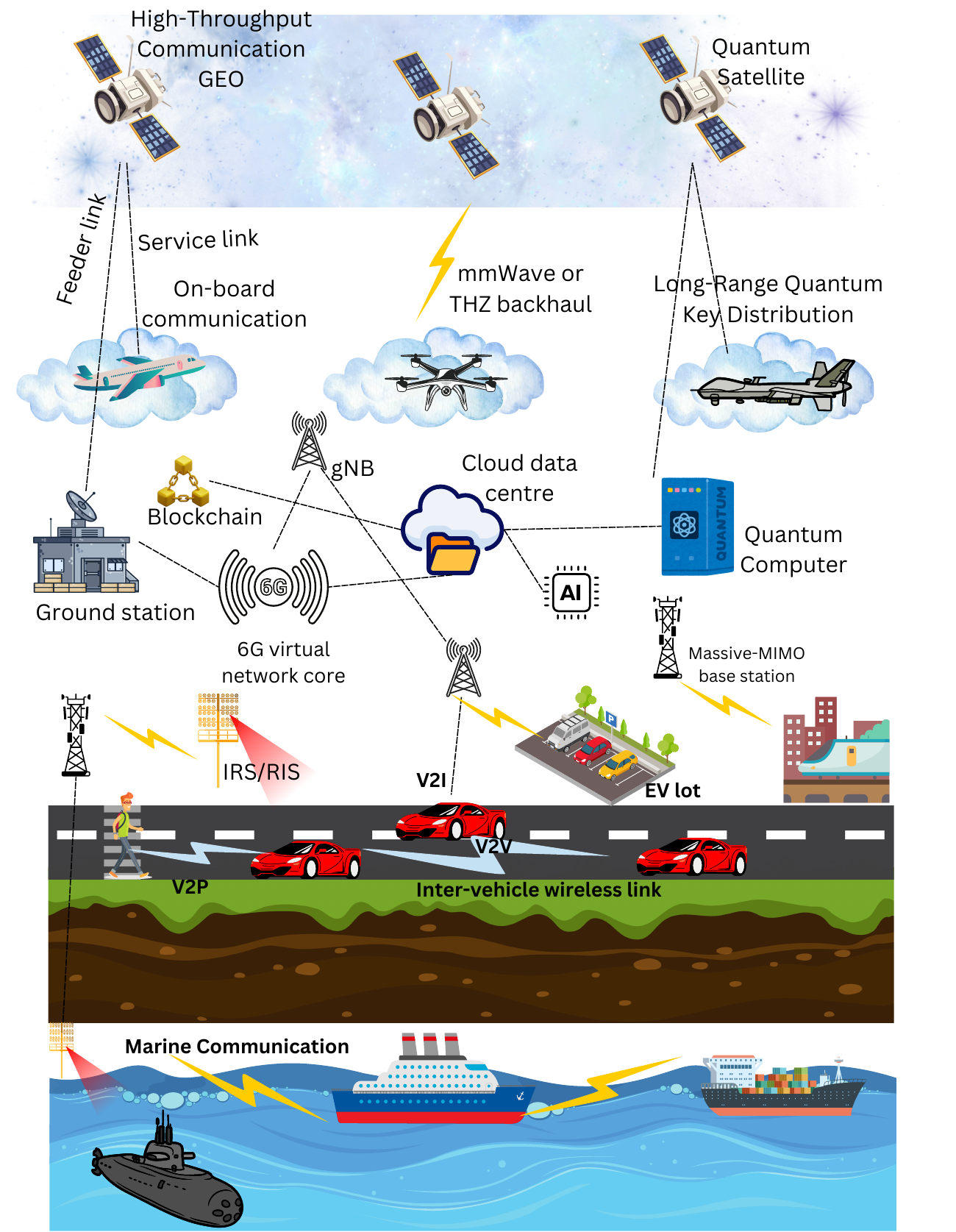}
    \caption{Vision of ground, sea, air and space 6G integrated network with ITS.}
    \label{fig:eco}
\end{figure}

\IEEEPARstart{A}{s} we approach the 2030s, the wireless communication landscape is poised for another revolutionary leap with the advent of sixth-generation (6G) technology~\cite{1}. Whilst fifth-generation (5G) networks are still in their deployment phase and their global adoption and coverage remain incomplete~\cite{2}, researchers and industry leaders are already envisioning the next frontier of connectivity that will seamlessly integrate communication across terrestrial, aerial, and space levels~\cite{2}.
6G represents the next generation of wireless communication technology, designed to surpass the capabilities of 5G in terms of speed, capacity, latency, and connectivity~\cite{6}. At its core, 6G employs a novel three-dimensional integrated network architecture that combines terrestrial, aerial, and satellite communications into a unified system~\cite{2,171}. This architecture consists of a terrestrial layer with ultra-dense networks of macro, micro, and nano cells, augmented with intelligent reflecting surfaces; an aerial layer utilizing high-altitude platforms and unmanned aerial vehicles to provide coverage and capacity in challenging environments; and a space layer comprising Low Earth Orbit (LEO) and Geostationary (GEO) satellites to ensure global coverage and connectivity~\cite{5}. These layers are interconnected through advanced backhaul and fronthaul networks, creating a seamless, three-dimensional coverage ecosystem~\cite{7} as shown in figure~\ref{fig:eco}.
Several groundbreaking features distinguish 6G. It boasts an Artificial Intelligence (AI)-native design, with AI embedded at every network layer, from the physical infrastructure to the application level, enabling autonomous network optimization, predictive resource allocation, and intelligent service provisioning~\cite{142,170}. Integrating quantum technologies promises ultra-secure communication channels through quantum key distribution and quantum-resistant cryptography~\cite{142}. Intelligent reflecting surfaces, comprising programmable metasurfaces, can dynamically control and optimize electromagnetic wave propagation, enhancing coverage, capacity, and energy efficiency~\cite{158}. Using terahertz~(THz) frequencies in the 0.1-10 THz range aims to achieve unprecedented data rates and support holographic communications~\cite{165}. Cell-free massive Multiple-Input Multiple-Output (MIMO) technology also provides uniform high-capacity coverage and mitigates cell-edge issues prevalent in traditional cellular networks~\cite{166}.

6G networks are designed to meet unprecedented performance metrics to support its ambitious goals. These include peak data rates of up to 1 terabit per second (Tbps), user-experience data rates of 1 gigabit per second (Gbps), spectrum efficiency 3 to 5 times higher than 5G, and network energy efficiency ten to hundred times better than its predecessor~\cite{166,167}. Furthermore, 6G aims to support an area traffic capacity of 1 Gbps/m\textsuperscript{2}, a connection density of 10 million devices per square kilometre ($10^7$ devices/km\textsuperscript{2}), latency below 100 microseconds, mobility up to 1000 kilometres per hour, and centimetre-level positioning accuracy~\cite{6,7,11}. These requirements are crucial for enabling advanced applications across various sectors, particularly in ITS.

6G's capabilities are set to revolutionize ITS by enabling ultra-reliable, low-latency communications, massive machine-type communications, enhanced mobile broadband, precise positioning, and AI-driven network optimization~\cite{6,169}. With latency below 100 microseconds, 6G enables real-time decision-making for autonomous vehicles, which is crucial for collision avoidance and platooning~\cite{168}. At highway speeds, this translates to a reaction distance of less than 3 millimetres, compared to ten centimetres with 5G~\cite{11}. The ability to support $10^7$ devices/km\textsuperscript{2} allows for dense networks of vehicles, roadside units, and sensors in urban environments, enabling comprehensive real-time traffic monitoring and management~\cite{7}. Peak data rates of 1 Tbps facilitate the real-time transmission of high-definition sensor data, supporting advanced perception systems in autonomous vehicles and allowing for the exchange of three-dimensional~(3D) environmental maps and high-resolution video streams among vehicles~\cite{169}. Centimetre-level positioning accuracy enhances navigation systems, enabling precise lane-level positioning for autonomous vehicles and more efficient traffic flow management~\cite{167}. The AI-native architecture allows for predictive traffic management, adaptive routing based on real-time data analysis, and proactive maintenance of transportation infrastructure~\cite{166}.

These advancements pave the way for transformative ITS applications such as fully autonomous transportation systems, seamless intermodal mobility, and intelligent traffic management that can adapt in real time to changing conditions~\cite{15}. However, the security implications of the increased connectivity and interoperability support of millions of loosely connected heterogeneous devices and vehicles present significant challenges, particularly in terms of security, privacy and trust~\cite{2}. The risks associated with security breaches in ITS could have severe consequences beyond financial loss or reputational damage, potentially impacting human safety. For instance, a cyber attack on an autonomous vehicle or any ITS component could lead to catastrophic accidents.

Realizing the transformational potential of 6G-powered ITS critically depends on establishing robust, multi-faceted security and trust between components~\cite{2}. Open challenges remain in balancing privacy with vast 6G-generated data to refine usability~\cite{126}. With billions of vehicles, devices, and infrastructure requiring seamless coordination, the complexity multiplies vulnerabilities that could endanger public safety if trust and security breaks~\cite{127}. Holistic trust and security mechanisms spanning technological, governance, and social dimensions are thus imperative but challenging to achieve. 

ITS integrates advanced sensors, communication modules, and data management systems to improve various aspects of transportation. For a more in-depth discussion of the ITS components and their implications on cyber security, refer to~\cite{16,17} and~\cite{25}.
\noindent Figure~\ref{fig:eco} illustrates 6G vision for ground, sea, air and space integrated network with ITS application.

This paper aims to answer the research questions highlighted in Table~\ref{tab:RQ}. The primary contributions of this research are summarized as follows:
\begin{enumerate}
    \item Comparison of the security of 5G-enabled ITS and 6G-ITS, analysing how the unique capabilities of 6G pose new security threats that must be addressed.
    \item Providing a taxonomy of the main security threats to 6G-ITS, including communication-based attacks, data-based attacks, device-based attacks, application-based attacks, privacy invasion and quantum attacks, and their impact on confidentiality, integrity, availability, and non-repudiation.
    \item Providing an in-depth analysis of the privacy risks in 6G-ITS and the impact of emerging privacy-preserving technologies, such as Federated Learning~(FL), differential privacy, and secret sharing schemes.
    \item Highlighting the key challenges of authentication and trust in 6G-ITS, especially in the context of the increased connectivity density and heterogeneous, decentralized nature of the networks. 
\end{enumerate}

\begin{table*}[]
    \centering
    \caption{Research Questions}
    
    \begin{tabular}{p{5cm} p{10cm}}
    \hline
        Research Questions (RQ) &  Discussion \\ \hline
        \textbf{RQ1:} In the context of 6G-ITS, how can multi-layered security strategies be designed to protect against sophisticated cyberattacks targeting communication networks, devices, and data analytics platforms? & This question addresses the critical aspect of security in 6G-ITS. It recognizes the complexity of potential cyber threats and the need for comprehensive, layered strategies to safeguard essential components of ITS. Addressing this question will provide insights into developing robust security frameworks that can adapt to the evolving threat landscape, ensuring the resilience of ITS infrastructure. \\
        \\
        \textbf{RQ2:} What mechanisms can be developed within 6G-ITS to ensure user privacy in the face of extensive data collection required for advanced sensing, automation, and communication systems? &  Privacy is a paramount concern in the deployment of ITS, especially with the advent of 6G technologies enabling unprecedented data exchange rates. This question underscores the challenge of balancing the benefits of enhanced data analytics capabilities with the imperative of protecting individual privacy. Solutions to this query are pivotal for fostering user trust and facilitating the wider acceptance of 6G-ITS.   \\
        \\
        \textbf{RQ3:}  How can trust in 6G-ITS be quantified and managed, particularly in systems involving decision-making and autonomous operations? & Quantifying and managing trust in 6G-ITS involves developing metrics to assess the reliability, security, and performance of AI/ML-driven technologies, alongside implementing dynamic assessment methods and adaptive protocols for ongoing monitoring and system improvements.\\
        \\
    \end{tabular}\label{tab:RQ}
\end{table*}

\begin{figure*}
    \centering
    \includegraphics[width=\textwidth]{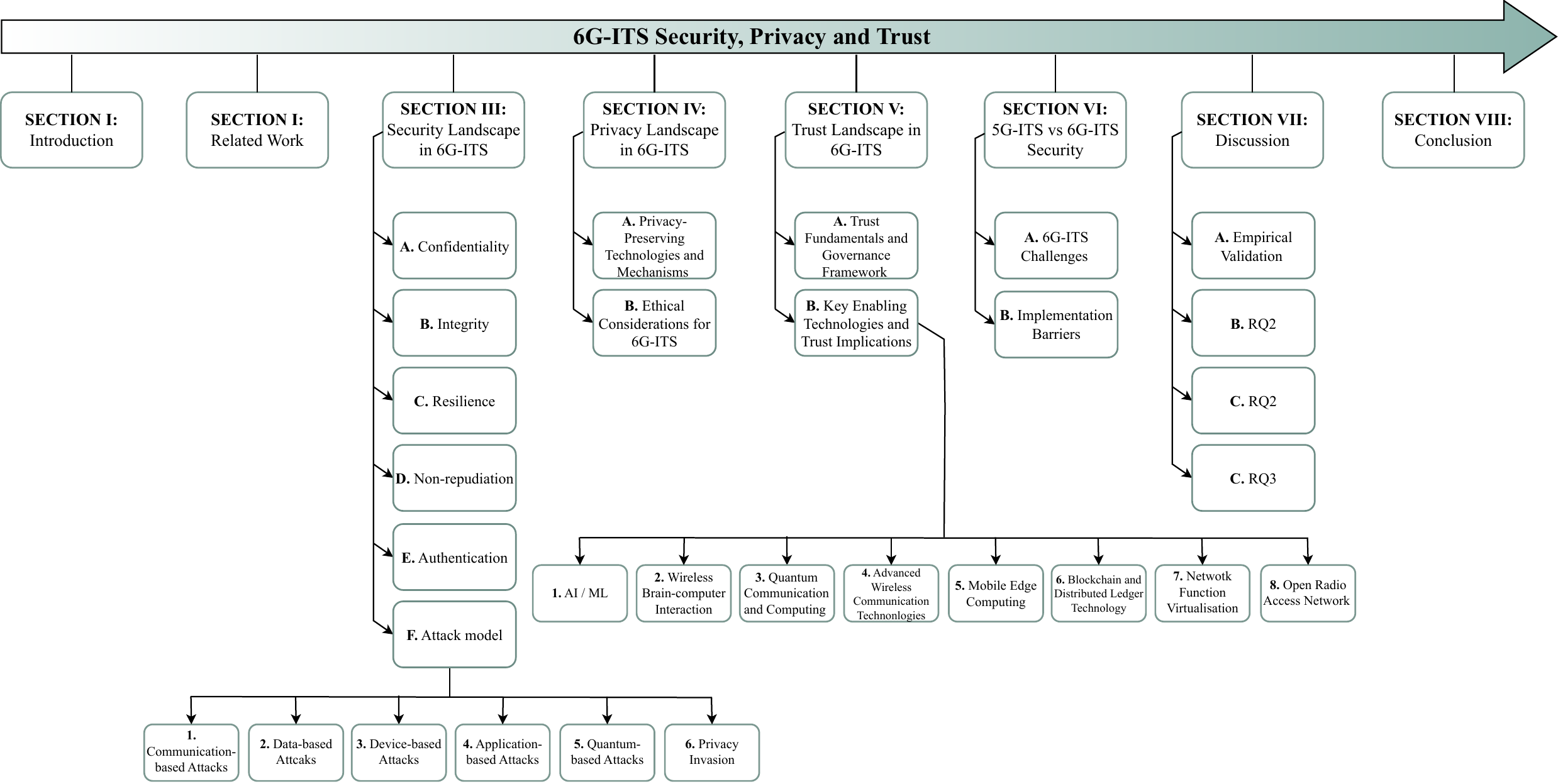}
    \caption{Structure and organization of this paper.}
    \label{fig:structure}
\end{figure*}

The remainder of this paper follows the structure illustrated in Figure~\ref{fig:structure}. Section~\ref{rwork} discusses existing works. Section~\ref{sec:security} examines the security landscape of 6G-ITS, where the attack model and security requirements are analyzed. Section \ref{sec:privacy} explains the privacy outlook of 6G-ITS, while Section~\ref{sec:trust} discusses the trust landscape in 6G-ITS. Section~\ref{sec:5Gvs6G} explores the differences between 5G-ITS and 6G-ITS. Section~\ref{sec:dis}~and~\ref{sec:conclusion} answer the research questions and conclude the paper, respectively. Table~\ref{symbols} contains the list of utilized acronyms.

{
\renewcommand{\arraystretch}{0.8}
\begin{table*}[!htb]
\caption{Main acronyms used in this paper.}
\label{symbols}
\centering
\resizebox{\linewidth}{!}{
\begin{tabular}{
    >{\arraybackslash}m{1.2cm}
    >{\arraybackslash}m{6cm}|
    >{\arraybackslash}m{1.2cm}
    >{\arraybackslash}m{6.18cm}
}
\textbf{Acronym} & \textbf{Definition} & \textbf{Acronym} & \textbf{Definition} \\ [0.3ex] 
\hline

3D & Three-Dimensional & ML & Machine Learning \\
3GPP & Third Generation Partnership Project & mmWave & Millimetre Wave \\
5G & Fifth Generation & NFV & Network Function Virtualization \\
6G & Sixth Generation & NOMA & Non-Orthogonal Multiple Access \\
6G-AITS & 6G Autonomous Intelligent Transportation Systems & OBUs & On-Board Units \\
6G-ITS & 6G-enabled Intelligent Transportation Systems & O-RAN & Open Radio Access Network \\
AI & Artificial Intelligence & P2BA & Privacy-preserving Protocol with Batch Authentication \\
AKA & Authentication and Key Agreement & PCA & Pilot Contamination Attacks \\
AMF & Access and Mobility Management Function & PKI & Public Key Infrastructure \\
APs & Access Points & PLS & Physical Layer Security \\
BCV & Brain-Controlled Vehicles & PQC & Post-Quantum Cryptography \\
C-FL & Consensus-driven Federated Learning & QKD & Quantum Key Distribution \\
CA & Central Authority & QoS & Quality of Service \\
CAV & Connected Autonomous Vehicle & RATs & Radio Access Technologies \\
DDoS & Distributed Denial of Service & RIC & Radio Intelligent Controller \\
DLT & Distributed Ledger Technology & RIS & Reconfigurable Intelligent Surface \\
dMIMO & Distributed Multiple-Input Multiple-Output & RSU & Road Side Unit \\
DP & Differential Privacy & SADC & Sensor Attack Detection Classification \\
EEG & Electroencephalography & SAGIN & Space-Air-Ground-Sea Integration Network \\
ETSI & European Telecommunications Standards Institute & SDN & Software Defined Networking \\
FL & Federated Learning & SDR & Software Defined Radio \\
GANs & Generative Adversarial Networks & SMC & Secure Multi-party Computation \\
Gbps & Gigabits per second & TA & Trusted Authority \\
GEO & Geostationary Earth Orbit & Tbps & Terabits per second \\
gNB & Next Generation NodeB & THz & Terahertz \\
IEEE & Institute of Electrical and Electronics Engineers & TLS & Transport Layer Security \\
IoT & Internet of Things & TPM & Trusted Platform Module \\
IoV & Internet of Vehicles & V2G & Vehicle-to-Grid \\
IRS & Intelligent Reflecting Surface & V2I & Vehicle-to-Infrastructure \\
ISAC & Integrated Sensing and Communication & V2N & Vehicle-to-Network \\
ITS & Intelligent Transportation Systems & V2P & Vehicle-to-Pedestrian \\
LEO & Low Earth Orbit & V2V & Vehicle-to-Vehicle \\
Li-Fi & Light Fidelity & V2X & Vehicle-to-Everything \\
Lidar & Light Detection and Ranging & VANET & Vehicular Ad Hoc Network \\
LSTM & Long Short-Term Memory & VLC & Visible Light Communication \\
MDI & Measurement-Device-Independent & VNF & Virtual Network Function \\
MEC & Mobile Edge Computing & VPN & Virtual Private Network \\
MIMO & Multiple-Input Multiple-Output & xApps & Extensible Applications \\
MitM & Man-in-the-Middle & ZTA & Zero Trust Architecture \\
\end{tabular}
}
\end{table*}
}

\section{RELATED WORK}\label{rwork}
The emergence of 6G technology sparked significant research interest in its potential applications, challenges, and implications for various sectors, including Intelligent Transportation Systems (ITS). The current state of research in 6G-ITS encompasses a wide range of themes, from privacy and security to trust management and emerging technologies.

Several studies outlined the vision and requirements for 6G networks. Jiang et al.~\cite{165} provide a comprehensive survey of 6G systems, discussing drivers, use cases, requirements, and enabling technologies. They predict an explosive growth in mobile traffic by 2030 and envision potential use cases and scenarios. Similarly, Zhang et al.~\cite{166} present a vision of 6G networks, describing usage scenarios and requirements for multi-terabytes per second and intelligent networks. They propose a large-dimensional and autonomous network architecture integrating space, air, ground, and underwater networks. Saad et al.~\cite{167} offer a holistic vision of 6G systems, arguing that 6G will be a convergence of technological trends driven by underlying services rather than mere exploration of higher frequency bands. They identify primary drivers of 6G systems and propose new service classes with target performance requirements.

Several researchers explored the application of 6G in ITS. Deng et al.~\cite{115} comprehensively review 6G autonomous intelligent transportation systems (6G-AITS), discussing the mechanisms, applications, and challenges. They emphasize the importance of maintaining a human-centric approach in the development of 6G-ITS. Jha et al.~\cite{168} investigate the potential of 6G in revolutionizing transportation systems, analysing the standards, technologies, and challenges associated with its implementation. They discuss novel applications such as autonomous driving, smart traffic management, and cooperative collision avoidance. Nguyen et al.~\cite{169} focus on the evolution of vehicular networks towards intelligent vehicular networks in 6G. They highlight why 5G is inadequate for specific vehicular applications and how 6G technologies can fill the gap. The work of Noor-A-Rahim et al.~\cite{26} shifts focus to enabling technologies for 6G-V2X communication, touching upon advancements in machine learning for vehicular networks. Similarly, Kirubasri et al.~\cite{27} survey the broader scope of 6G vehicular technology, its applications, and the challenges it presents, indicating a paradigm shift from smart to intelligent systems.

Security and privacy concerns in 6G-ITS are a significant focus of research. Osorio et al.~\cite{24} provide a comprehensive survey on security and privacy in 6G-enabled Internet of Vehicles (IoV), covering the evolution of V2X communication towards IoV and highlighting security frameworks and privacy concerns. Wang et al.~\cite{5} discuss new areas and challenges in security and privacy for 6G networks, including real-time intelligent edge computing, distributed artificial intelligence, intelligent radio, and 3D intercoms. They also explore potential use cases and emerging technologies in each area. Nguyen et al.~\cite{68} offer a systematic overview of security and privacy issues based on prospective technologies for 6G in the physical, connection, and service layers. They highlight new threat vectors from new radio technologies and discuss promising techniques to mitigate the magnitude of attacks and breaches of personal data. The research in~\cite{25} focuses on the current security landscape of ITS, particularly considering the integration of the Internet of Things~(IoT). It thoroughly examines ITS's security posture, including attacker models and potential vulnerabilities. Complementing this, a survey presented in~\cite{16} categorizes the security challenges in ITS, identifying major attack types, such as DDOS and session hijacking. 

Trust management in 6G networks has emerged as a crucial research area. Ziegler et al.~\cite{127} discuss the evolution of the 5G security paradigm and explore relevant security technology enablers for 6G, including automated software creation, privacy-preserving technologies, and quantum-safe security. Wang et al.~\cite{123} introduce a new trust framework called SIX-Trust, made up of three layers that focus on sustainable trust, infrastructure trust, and xenogenesis trust. They demonstrate how these technologies can enhance the trust and security of 6G networks. Veith et al.~\cite{125} provide an overview of trust anchor technologies for 6G, describing the requirements for an end-to-end trust building framework, and discussing the concept of trust in mobile communications systems.

Several studies explored emerging technologies that could enable 6G-ITS. Akyildiz et al.~\cite{170} discuss transformative solutions expected to drive the surge for accommodating a rapidly growing number of intelligent devices and services in 6G and beyond. They highlight technologies such as THz band communications, intelligent communication environments, and pervasive artificial intelligence. Giordani et al.~\cite{171} discuss technologies that will evolve wireless networks toward 6G, providing a full-stack, system-level perspective on 6G scenarios and requirements. They focus on technologies that can satisfy these requirements by improving the 5G design or introducing completely new communication paradigms. Yang et al.~\cite{172} present an overview of promising techniques evolving to 6G, including physical-layer transmission techniques, network designs, security approaches, and testbed developments.
\noindent A comparison of previous related surveys and how our work fills the existing gap is presented in Table~\ref{tab:survey}.

\definecolor{verylightgray}{gray}{0.95}
\definecolor{lightgreen}{rgb}{0.88, 1, 0.88}

{
\renewcommand{\arraystretch}{1.4}
\begin{table*}[!htb]
\caption{A comparison of related surveys in the literature.} Annotations: "\checkmark" indicates covered, and "\text{\sffamily X}" indicates not covered
\label{tab:survey}
\begin{center}
\resizebox{\linewidth}{!}{
\begin{tabular}{|>{\centering\arraybackslash}m{0.8cm}|>{\centering\arraybackslash}m{0.7cm}>{\centering\arraybackslash}m{0.7cm}>{\centering\arraybackslash}m{0.9cm}>{\centering\arraybackslash}m{0.9cm}>{\centering\arraybackslash}m{0.9cm}|>{\arraybackslash}m{4.95cm}|>{\arraybackslash}m{5cm}|}
\hline
\rowcolor{verylightgray}
\textbf{Ref} & \textbf{6G} & \textbf{ITS} & \textbf{Security} & \textbf{Privacy} & \textbf{Trust} & \multicolumn{1}{c|}{\textbf{Focus}} & \multicolumn{1}{c|}{\textbf{Limitation}} \\
\hline 
\cite{165} & \checkmark & \text{\sffamily X} & \text{\sffamily X} & \text{\sffamily X} & \text{\sffamily X} & Comprehensive survey of 6G systems & Limited focus on ITS and security \\
\hline
\cite{166} & \checkmark & \checkmark & \text{\sffamily X} & \text{\sffamily X} & \text{\sffamily X} & Large-dimensional 6G network architecture & Lacks security and privacy considerations \\
\hline
\cite{167} & \checkmark & \text{\sffamily X} & \text{\sffamily X} & \text{\sffamily X} & \text{\sffamily X} & Holistic vision of 6G systems & Does not address ITS or security challenges \\
\hline
\cite{115} & \checkmark & \checkmark & \text{\sffamily X} & \text{\sffamily X} & \text{\sffamily X} & 6G autonomous ITS & Limited security, privacy, and trust focus \\
\hline
\cite{168} & \checkmark & \checkmark & \text{\sffamily X} & \text{\sffamily X} & \text{\sffamily X} & 6G in revolutionizing transportation & Lacks in-depth security analysis \\
\hline
\cite{169} & \checkmark & \checkmark & \text{\sffamily X} & \text{\sffamily X} & \text{\sffamily X} & Intelligent vehicular networks in 6G & Inadequate security and privacy coverage \\
\hline
\cite{26} & \checkmark & \checkmark & \text{\sffamily X} & \text{\sffamily X} & \text{\sffamily X} & Enabling technologies for 6G-V2X & Limited security and trust considerations \\
\hline
\cite{27} & \checkmark & \checkmark & \text{\sffamily X} & \text{\sffamily X} & \text{\sffamily X} & Broader scope of 6G vehicular technology & Lacks detailed security analysis \\
\hline
\cite{24} & \checkmark & \checkmark & \checkmark & \checkmark & \text{\sffamily X} & Security and privacy in 6G-IoV & Limited trust management focus \\
\hline
\cite{5} & \checkmark & \text{\sffamily X} & \checkmark & \checkmark & \text{\sffamily X} & Security and privacy in 6G networks & Not specifically focused on ITS \\
\hline
\cite{68} & \checkmark & \text{\sffamily X} & \checkmark & \checkmark & \text{\sffamily X} & Security and privacy in 6G technologies & Lacks ITS-specific considerations \\
\hline
\cite{25} & \text{\sffamily X} & \checkmark & \checkmark & \text{\sffamily X} & \text{\sffamily X} & Security landscape of ITS & Limited focus on 6G and privacy \\
\hline
\cite{16} & \text{\sffamily X} & \checkmark & \checkmark & \text{\sffamily X} & \text{\sffamily X} & Security challenges in ITS & Does not address 6G specifics \\
\hline
\cite{127} & \checkmark & \text{\sffamily X} & \checkmark & \checkmark & \checkmark & 6G security paradigm evolution & Not focused on ITS applications \\
\hline
\cite{123} & \checkmark & \text{\sffamily X} & \text{\sffamily X} & \text{\sffamily X} & \checkmark & SIX-Trust framework for 6G & Limited security and privacy coverage \\
\hline
\cite{125} & \checkmark & \text{\sffamily X} & \text{\sffamily X} & \text{\sffamily X} & \checkmark & Trust anchor technologies for 6G & Lacks ITS-specific considerations \\
\hline
\cite{170} & \checkmark & \text{\sffamily X} & \text{\sffamily X} & \text{\sffamily X} & \text{\sffamily X} & Transformative solutions for 6G & Limited security, privacy, trust in ITS \\
\hline
\cite{171} & \checkmark & \text{\sffamily X} & \text{\sffamily X} & \text{\sffamily X} & \text{\sffamily X} & Technologies for 6G networks & Lacks security and ITS considerations \\
\hline
\cite{172} & \checkmark & \text{\sffamily X} & \checkmark & \text{\sffamily X} & \text{\sffamily X} & Promising techniques for 6G & Limited privacy, trust, ITS applications \\
\hline
\rowcolor{lightgreen}

Our Work & \checkmark & \checkmark & \checkmark & \checkmark & \checkmark & Comprehensive security, privacy, and trust in 6G-ITS & Limited focus on practical implementation challenges and experimental validation  \\
\hline
\end{tabular}%
}
\end{center}

\end{table*}
}

\section{SECURITY LANDSCAPE IN 6G-ITS}\label{sec:security}
\begin{figure*}
    \centering
    \includegraphics[width=\textwidth, trim=20 5 40 5, clip]{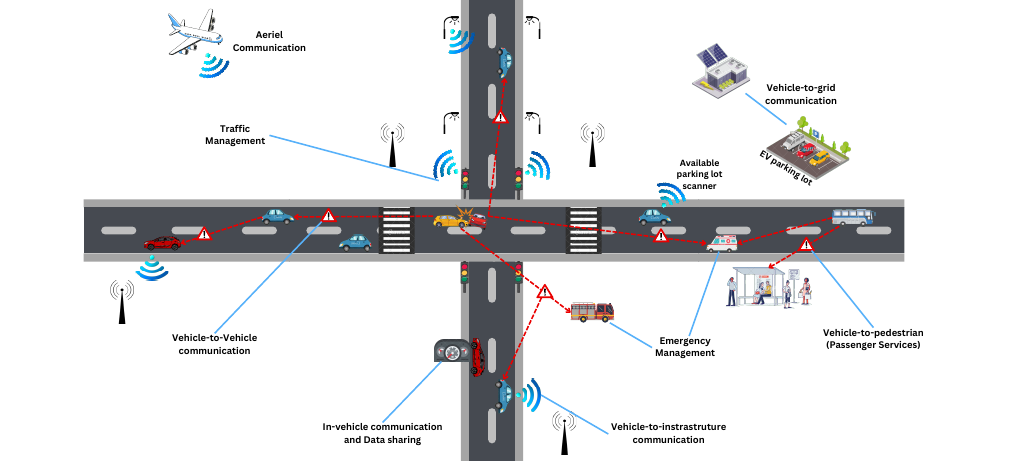} 
    \caption{High-level overview of ITS architecture.}
    \label{fig:ITS}
\end{figure*}

Advancements in 6G technology introduce unprecedented challenges to the security landscape of ITS. With its ultra-high bandwidth, massive connectivity, and AI/ML integration at the network core, 6G fundamentally alters the nature and scale of security threats. The 6G-ITS security landscape is complex, connecting diverse components like vehicles, drones, and several IoT devices. This heterogeneity and high mobility complicate achieving both security and interoperability~\cite{16}, necessitating advanced authentication and security measures. This section examines how 6G impacts confidentiality, integrity, availability, authentication, and non-repudiation in ITS, exploring evolving security requirements and new attack models. Figure~\ref{fig:ITS} illustrates a high-level overview of the ITS architecture based on an accident emergence scenario.


\subsection{Confidentiality}
Ensuring security and confidentiality in a 6G-ITS is essential. 
6G's massive connectivity significantly increases the attack surface for data breaches. Confidentiality attacks could include attacks on the communication systems that connect vehicles, infrastructure, and control centres or on the data analytics and management tools that process and analyse the data collected by the sensors and cameras. To mitigate these risks, it is important to use secure protocols and encryption to protect communication and robust security measures, such as firewalls, intrusion detection systems, and incident response plans, to protect the system from cyberattacks~\cite{16}. Data breaches pose a significant risk to the security of ITS, particularly as these systems collect sensitive data such as traffic patterns, weather conditions, and pedestrian movements, which may include personally identifiable information. Unlike 5G, 6G's integration of AI/ML at the network core introduces new data processing and analysis vulnerabilities, such as adversarial examples and privacy leaks. With 6G enabling more sophisticated AI/ML integrations, enhanced data privacy measures are paramount to protect individual confidentiality. For instance, the trial of AI-powered speed cameras in the UK raised privacy concerns due to their invasive imaging capabilities~\cite{164}. Maintaining public trust will hinge on addressing these confidentiality issues as mobile technology evolves. It is important to use proper data encryption and management practices to protect this data and implement access controls to ensure that only authorized personnel can view or use it~\cite{63}.

Physical attacks on the system's infrastructure include sensors, cameras, and communication devices. It is important to secure the physical infrastructure of the ITS, for example, by using tamper-proof housing and cameras that can detect and respond to intrusions and ensure that critical infrastructure is well-protected and that access is strictly controlled~\cite{64}.


\subsection{Integrity}
One of the main concerns for security integrity in an ITS is maintaining the consistency and accuracy of the data being transmitted, processed and stored. The sheer volume of data in 6G networks, potentially reaching zettabytes per year, exponentially increases the challenge of maintaining data integrity. To maintain integrity, the data should be protected from unauthorized changes, deletions, or insertions during the transmission, processing and storage. Tampering detection and prevention mechanisms such as digital signature, authentication, and encryption should be implemented to ensure the integrity of the data. Another aspect to consider is the integrity of the system's components and communication. The ITS components, such as sensors, cameras, communication devices and other devices, should be secured and protected from physical tampering and unauthorized access. This includes the control system responsible for decision-making, coordination, and management of the system's resources~\cite{67}.

Additionally, the integrity of the system's software, including the Operating System and the Application Software, must be maintained. Regular updates and patch management should be done to address the known vulnerabilities and to keep the system free from malware and other malicious software. Furthermore, the integrity of the system's decision-making process and algorithms should be maintained, particularly when machine learning and AI techniques are used. This requires the validation and testing of the models to ensure their robustness and integrity. The security integrity of a 6G-ITS requires a holistic approach involving multiple layers of security controls and mechanisms to be implemented and maintained to protect the system and its data from unauthorized access, tampering, and malicious activities~\cite{16}.

\subsection{Resilience}
The ability to keep the system running in the face of cyberattacks or other disruptive events is one of the primary problems related to security resilience in an ITS. With 6G's sub-millisecond latency requirements, even brief disruptions can have severe consequences, making resilience a critical concern. This involves defending the system against ransomware, DDoS (Distributed Denial of Service) assaults, and other malicious software that can render the system inoperable. Implementing security measures like firewalls, intrusion detection and prevention systems, and incident response strategies is crucial to reducing these risks. Another challenge for security availability in an ITS is ensuring the resilience and fault tolerance of the system. This includes protecting the system from hardware or software failures, power outages, or other disruptions that could cause the system to become unavailable~\cite{16}. The increased reliance on edge computing in 6G networks introduces new points of failure that must be addressed to maintain system availability.

To mitigate these risks, it is important to implement redundant and diverse systems and network designs and to develop detailed disaster recovery plans that can be activated in the event of an incident. Moreover, the integration of emerging technologies, such as AI and ML, raises additional concerns for security and availability, including ensuring the robustness and transparency of the system's decision-making process and maintaining system availability. Ensuring security availability in a 6G-ITS requires a combination of technical and organizational measures to protect the system from a wide range of threats and disruptions and to ensure that the system can continue to function even in the event of an incident or failure~\cite{30}.

\subsection{Non-repudiation}
Ensuring security non-repudiation in a 6G-ITS is an important aspect of the system's overall security, as it helps to ensure that the authenticity and integrity of the data and communications within the system can be verified and that any actions taken by the system can be traced to their originator~\cite{30}.

One of the main challenges for non-repudiation in an ITS is ensuring that the data and communications are properly authenticated and that the sender’s identity can be verified. This can be achieved using techniques such as digital signatures, certificates, and Public Key Infrastructure~(PKI) to authenticate the sender and the data and tamper-proof hardware or secure communication protocols to protect the data in transit~\cite{33}. However, the quantum computing capabilities expected in the 6G era pose a significant threat to traditional PKI systems, necessitating the development of quantum-resistant cryptographic techniques. 

Non-repudiation in an ITS is maintaining a tamper-proof log of all actions and events within the system, although this will result in high computational costs. However, this can be achieved by using digital signatures, digital time-stamping, and other technologies that 6G promises to ensure that actions integrated computing and events can be traced back to their originator and by using secure data storage and archiving practices to preserve the logs for future reference~\cite{33,68}.

Non-repudiation is also essential for the liability and accountability aspect of the system. In case of incidents and accidents, it is vital to trace the decisions and actions that led to the incident and hold the responsible parties accountable. This becomes particularly complex in 6G-ITS, where decision-making may involve AI and machine learning algorithms, raising questions about accountability in automated systems.

\subsection{Authentication}
Authentication is a crucial aspect of security in a 6G-ITS, as it helps to ensure that only authorized entities can access and interact with the system. The heterogeneity and massive scale of devices in 6G networks make authentication a particularly complex challenge. The ITS include various types of entities, such as vehicles, infrastructure, control centres and other devices that must be authenticated before accessing the system~\cite{69}. It is important to note that authentication in a 6G-ITS is bidirectional, meaning that user devices must authenticate themselves to the system, and the 6G infrastructure must also authenticate itself to the user devices. While this bidirectional authentication helps prevent attacks realized via fake base stations and ensures that users connect to legitimate network infrastructure, it also introduces new security vulnerabilities to replay attacks, forgery attacks and man-in-the-middle attacks~\cite{1}. The scalability and flexibility of the system are among the primary difficulties in authentication in an ITS~\cite{7}. It gets increasingly challenging to manage and protect the system and ensure that the appropriate entities are given access at the appropriate level as the number of entities engaging with the system grows.

Using secure and adaptable authentication techniques, such as multi-factor authentication, digital certificates, and Public-Key Infrastructure (PKI), that can expand to handle many entities without sacrificing security is crucial to overcoming this difficulty. Authentication in an ITS maintains the privacy and security of the personal information used for authentication~\cite{70}. Personal information, such as biometric data, can be sensitive and should be protected from unauthorized access and misuse. It is important to address this challenge using secure data storage and encryption and to implement strict access controls and audit trails. Authentication drawbacks in an ITS are the integration of IoT devices that may have limited resources and computational capabilities, and it is challenging to implement robust and secure authentication methods on such devices. Integrating emerging technologies such as AI and ML and other 6G enabling technologies poses additional challenges, such as the robustness of the authentication methods against advanced attacks and the transparency and fairness of the decision-making process~\cite{71}. State of the art in authentication schemes and the challenges faced in 6G-ITS will further be analyzed in Section~\ref{sec:dis}.

\subsection{Attack model for 6G-ITS}\label{sec:attack}
\begin{figure*}[!ht]
    \centering
   \includegraphics[width=\textwidth, keepaspectratio]{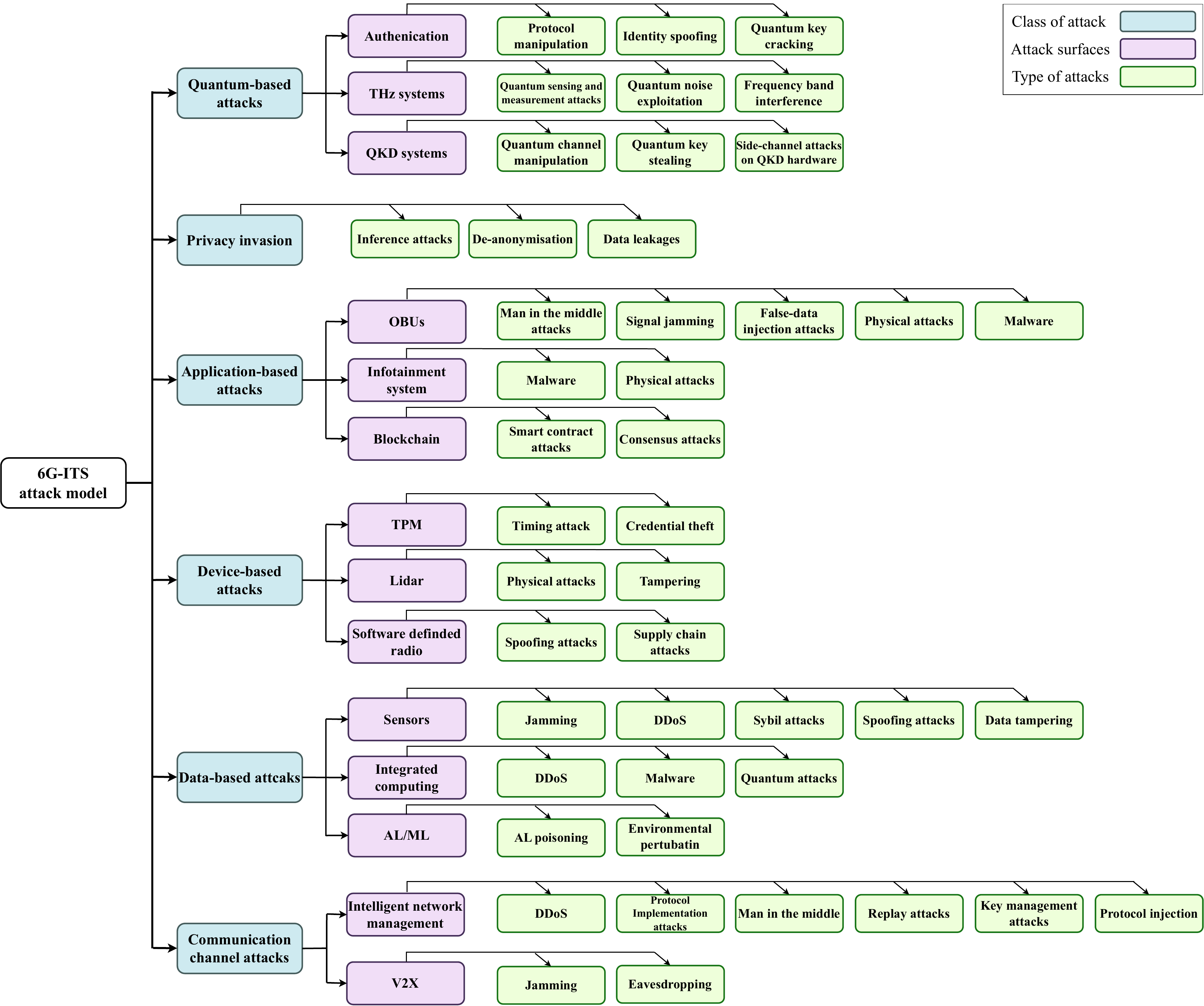}
    \caption{Attack model of 6G-ITS.}
    \label{fig:attack}
\end{figure*}

In recent years, there has been an increase in cyberattack publications on several components of ITS, particularly VANET and Connected Autonomous Vehicles (CAVs). As identified by Guo et al.~\cite{33}, security threats in the 6G mobile network are classified into five categories based on the proposed Space-Air-Ground-Sea Integration Network~(SASGIN) architecture. These are communication-based attacks, data-based attacks, Device-based attacks, application-based attacks, and quantum-based attacks. See Figure~\ref{fig:attack} for an illustrated attack model on 6G-ITS.

\subsubsection{Communication-based Attacks}
Communication-based attacks in a 6G-ITS target the communication channels and protocols used by the system to transmit data and control information. These types of attacks primarily exploit vulnerabilities in V2X infrastructure and intelligent network management systems. 

The V2X infrastructure represents a critical attack surface in 6G-ITS, encompassing communication channels between vehicles (V2V), vehicles to infrastructure (V2I), vehicles to pedestrians (V2P), and vehicles to network (V2N). Operating at unprecedented data rates of up to 1 Tbps and ultra-low latency in the sub-millisecond range, V2X in 6G introduces new vulnerabilities through its ultra-dense network deployment, complex heterogeneous architecture, dynamic network topology, and stringent real-time requirements for safety-critical applications. These characteristics create numerous potential entry points for malicious actors seeking to compromise system integrity.

Similarly, Intelligent Network Management Systems form a sophisticated attack surface responsible for network resource orchestration, Quality of Service (QoS) management, network slicing, virtualization, and dynamic spectrum allocation. The increased complexity and autonomy of these systems in 6G create new attack vectors through their decision-making algorithms and control interfaces. The ability of these systems to dynamically reconfigure network resources and manage multiple virtual network slices simultaneously presents unique security challenges in the 6G context.
\noindent A common attack model for communication-based attacks in a 6G-ITS would include the following:
\begin{itemize}
    \item Distributed Denial of Service (DDoS) attacks: These attacks target the availability of the communication channels by overwhelming the system with a large amount of traffic, making it unable to process legitimate requests~\cite{72}.
    \item Man-in-the-Middle (MitM) attacks intercept and manipulate the communication between different entities in the ITS, such as vehicles, infrastructure, and control centres, to steal or corrupt data.
    \item Key management attacks: These attacks target the cryptographic infrastructure underlying secure communications~\cite{72}.
    \item Replay attacks: These attacks involve recording and replaying valid communication messages to impersonate a legitimate entity and gain unauthorized access to the system.
    \item Protocol injection attacks: These attacks involve injecting malicious code into the communication protocol to cause a denial of service or gain unauthorized access to the system~\cite{62}.
    \item Protocol implementation attacks: These attacks exploit vulnerabilities in communication protocol implementations to compromise system integrity.
    \item Eavesdropping: Targets the expanded communication channels in 6G networks to intercept sensitive transmissions between devices, vehicles, and infrastructure. The high bandwidth and complex beam-forming techniques of 6G create new opportunities for covert data capture.
    \item Jamming: Introduces precise, targeted interference to disrupt wireless communications. In 6G networks, these attacks become more sophisticated due to the higher frequency bands and directional communications, allowing attackers to target specific network segments while evading detection. A notable example of such attacks in the UK is relay theft, particularly affecting Land Rover vehicles. In these cases, attackers use key fob jamming devices to trick the vehicle into believing the key is nearby, facilitating unauthorized access and theft~\cite{173}.
\end{itemize}

To mitigate these risks, it is important to implement security measures such as using secure communication protocols, such as quantum-based key distribution, implementing secure key management, using techniques like frequency hopping, and using secure methods for data encryption and message integrity checks. Additionally, implementing robust intrusion detection and response systems to quickly identify and mitigate communication-based attacks is crucial, as detailed in~\cite{63}.

\subsubsection{Data-based Attacks}
Data processing in 6G-ITS represents a multi-layered ecosystem where artificial intelligence, distributed computing, and sensor networks converge to enable unprecedented levels of automated decision-making and system optimization. At the core of this ecosystem, Artificial Intelligence and Machine Learning systems serve as the cognitive engine of 6G-ITS, processing massive amounts of data through sophisticated training pipelines and inference engines. The integration of AI/ML directly at the network core, a distinctive feature of 6G networks, fundamentally transforms the security landscape. While these systems enable crucial functionalities such as autonomous driving, predictive traffic management, and dynamic network optimization, their deep integration also expands the potential attack surface significantly. The complexity of distributed learning frameworks and AI-driven decision-making systems makes them particularly vulnerable to sophisticated attacks that could compromise the entire ITS infrastructure.

Supporting these AI systems, the integrated computing infrastructure forms a crucial layer that bridges the gap between raw data collection and intelligent decision-making. This infrastructure leverages a sophisticated combination of edge and cloud computing resources to meet the demanding computational requirements of 6G-ITS. The distributed nature of this computing fabric, while essential for achieving the ultra-low latency demands of 6G applications, introduces numerous potential points of compromise. Edge computing nodes, positioned closer to data sources, must process time-critical information within microseconds while maintaining security integrity. Meanwhile, cloud data centres handle more complex, resource-intensive computations, creating a dynamic computational environment where security must be maintained across multiple processing layers and resource allocation boundaries.

At the foundation of this data processing hierarchy lies the sensor infrastructure, which serves as the primary interface between the physical and digital domains of 6G-ITS. This sophisticated network of sensors encompasses environmental monitoring systems, vehicle-mounted devices, infrastructure sensors, and high-precision positioning systems, all operating at unprecedented data collection rates. The density and sophistication of these sensor networks in 6G environments create a vast attack surface where physical and cyber threats converge. Each sensor node not only collects critical data but also participates in the broader network fabric, making the security of these devices paramount to system integrity. The interconnected nature of these sensors, combined with their real-time data transmission requirements and the sheer volume of information they process, creates complex attack vectors that could potentially compromise both local and system-wide operations.
\noindent Examples of typical data-based attacks in a 6G-ITS include the following:
\begin{itemize}
    \item AI/ML poisoning: These attacks manipulate training data or introduce adversarial examples to compromise model behaviour, affecting critical functions such as autonomous vehicle control and traffic management. 
    \item Environmental perturbation: These attacks introduce engineered disturbances to affect sensor readings and data collection, exploiting the physical-digital interface of 6G-ITS systems.
    \item Data tampering: These attacks target the integrity of the data by modifying the data without authorization to disrupt the system’s normal functioning or gain unauthorized access to sensitive information.
    \item Data injection attacks: These attacks involve injecting false or malicious data into the system to disrupt the system’s normal functioning or gain unauthorized access to sensitive information~\cite{73}.
    \item Quantum attacks: Quantum attacks represent an emerging threat vector that exploits the quantum computing capabilities integrated into 6G networks. These attacks target both classical and quantum cryptographic systems used for data protection, potentially compromising the confidentiality of sensitive transportation data. The integration of quantum key distribution (QKD) and quantum-safe cryptography in 6G networks introduces new attack surfaces that must be carefully considered in security frameworks~\cite{40}.
    \item Sybil attacks: This attack creates multiple fake identities to manipulate data aggregation and decision-making, which is particularly dangerous in V2X communications. 
    \item Spoofing attacks: these attacks focus on compromising data authenticity and integrity, potentially affecting critical system operations through unauthorized modifications of sensor readings, vehicle telemetry, or traffic pattern data~\cite{73}.
    \item Malware: Malware in 6G-ITS environments has evolved to target sophisticated data processing pipelines that handle massive amounts of sensor and operational data. These attacks can propagate across different processing layers, from edge devices to cloud infrastructure, potentially compromising entire data processing chains. Modern malware in 6G contexts often employs AI-driven techniques to evade detection and adapt to defensive measures, making them particularly challenging to identify and mitigate~\cite{1}. 
\end{itemize}

Recent studies outline several targeted mitigation strategies to address data-based attacks in 6G-ITS. AI/ML poisoning is mitigated through pre-filtering and anomaly detection techniques that detect adversarial input before the model is ingested~\cite{s24020338}. Sensor-level threats such as environmental perturbation and spoofing are mitigated by real-time fusion and classification frameworks like Sensor Attack Detection Classification (SADC)~\cite{10323101}. To preserve data integrity, blockchain-enabled logging mechanisms offer tamper-proof verification in distributed ITS environments~\cite{10376064}. Quantum safe architectures combine post-quantum cryptography with QKD to defend against quantum-enabled attacks~\cite{math13132074, 10105165}. Sybil and spoofing attacks are addressed through trust frameworks using proof-of-location schemes for authenticated identity validation~\cite{s24248140}. AI-driven intrusion detection, often federated and explainable, strengthens defenses against malware and injection attacks at the edge and cloud layers~\cite{s25030854}.

\subsubsection{Device-based Attacks}
The device-level attack surface in 6G-ITS encompasses a sophisticated ecosystem of interconnected hardware components, each contributing to the system's enhanced capabilities while introducing unique security challenges. At the communication layer, Software Defined Radio (SDR) systems form the backbone of programmable wireless communications, representing a fundamental shift from traditional fixed-function radio hardware. These systems offer unprecedented flexibility through reconfigurable radio frequency interfaces and dynamic protocol implementations, but this very adaptability creates expanded attack surfaces. In 6G networks, where SDRs must handle complex waveforms and ultra-high-frequency communications, the increased sophistication of digital signal processing components and software control interfaces introduces new vulnerabilities that attackers could exploit to compromise system integrity.

Environmental perception in 6G-ITS relies heavily on Light Detection and Ranging (Lidar) systems, which serve as critical sensors for autonomous vehicle operation and infrastructure monitoring. These high-precision systems integrate sophisticated optical sensors with advanced signal processing units to enable real-time environmental mapping and object detection. The deployment of next-generation Lidar systems in 6G environments, characterized by enhanced resolution and expanded range capabilities, introduces complex attack surfaces at both the hardware and software levels. The intricate interplay between optical sensors, signal processing algorithms, and object detection systems creates multiple potential entry points for adversaries seeking to manipulate environmental perception data.

Securing these advanced hardware components falls largely to Trusted Platform Module (TPM) systems, which serve as hardware-based security anchors in the 6G-ITS infrastructure. While designed specifically for security, these modules must manage an increasingly complex set of responsibilities in 6G environments, from securing ultra-high-speed communications to supporting quantum-safe operations. The expanded role of TPMs in protecting key storage, executing cryptographic operations, maintaining secure boot processes, and providing attestation services creates a critical security foundation. However, the increased complexity of these security operations, coupled with the need to support both classical and quantum-safe cryptographic protocols, introduces new challenges in maintaining the integrity of these fundamental security components. 
\noindent The following are some of the attacks in this category:
\begin{itemize}
    \item Spoofing: Exploit programmable radio interfaces to forge device identities and manipulate wireless signals in 6G communications. These attacks take advantage of SDR's increased flexibility to impersonate legitimate network nodes or vehicle systems.
    \item Physical attacks: These attacks target the availability of the device by physically destroying it, making it unable to function or communicate with other entities in the ITS~\cite{30}.
    \item   Supply chain attacks: Target device integrity during manufacturing and distribution, potentially introducing hardware trojans or compromised firmware. The complexity of 6G devices, particularly those with quantum-safe components, creates multiple insertion points for malicious modifications~\cite{60}.
    \item Timing attacks: Exploit the ultra-precise synchronization requirements of 6G networks to disrupt system operations. These attacks are particularly effective against secure protocols and TPM operations where timing accuracy is crucial.
\end{itemize}
Recent works demonstrate an effective mitigation strategy towards these attacks. Integrated Sensing and Communication (ISAC) links GPS with radio fingerprints to spot SDR-based spoofing with high accuracy~\cite{10118852}. Hardware integrity is reinforced by machine-learning-enhanced Physical Unclonable Functions, which reach 99\% authentication success for on-board units~\cite{10.1145/3686625.3686636}. Supply chain risks are mitigated through post-quantum schemes, such as CRYSTALS-Kyber and CRYSTALS-Dilithium, which future-proof key exchanges and firmware signing~\cite{10105165}. Constant-time cryptographic routines combined with period-based randomization (PerRand) seal off micro-architectural timing channels while still meeting the microsecond-level synchronization demanded by safety-critical 6G applications~\cite{Berardi2023TSN}. These integrated approaches demonstrate robust attack mitigation effectiveness while preserving the microsecond-level timing guarantees essential for safety-critical 6G transportation applications.
\subsubsection{Application-based Attacks}
The application layer in 6G-ITS encompasses interconnected systems bridging user interactions, vehicle operations, and network services. On-board units (OBUs) manage the interface between vehicles and 6G infrastructure through sophisticated communication modules, processing units, and storage systems. Their integration with vehicle systems creates complex security considerations unique to 6G's ultra-high-speed requirements.

Vehicle infotainment systems have evolved into sophisticated computing platforms, serving as gateways between passengers and the transportation network. The expanded connectivity options and processing capabilities in 6G networks will make these systems particularly vulnerable to security breaches through their user interfaces and network connections.

Blockchain technology secures 6G-ITS operations through consensus protocols, smart contracts, and cryptographic implementations. While enhancing security through distributed trust, these systems introduce new complexities in managing consensus at 6G speeds across mobile nodes. The interplay between OBUs, infotainment systems, and blockchain networks demands security considerations spanning multiple technological domains.
\noindent The following are some of the application-based attacks:
\begin{itemize}
    \item Smart contract attacks: Target automated agreement systems in blockchain platforms, exploiting vulnerabilities in contract logic or execution. In 6G-ITS, these attacks can compromise automated vehicle services, charging systems, and traffic management contracts operating at ultra-high speeds~\cite{74}.
    \item Consensus attacks: Manipulate the decision-making processes of the blockchain network by exploiting the distributed nature of 6G networks. These attacks attempt to gain control over network validation mechanisms, potentially affecting vehicle coordination and service verification.
    \item False-Data Injection: Introduce fabricated or manipulated data into application processes, particularly targeting sensor data processing applications and traffic management systems in 6G-ITS~\cite{51}.
    \item Evasion attacks: These attacks target the AI and ML models used in the ITS by creating inputs that are specifically designed to evade the model's detection capabilities.
    \item Model-stealing attacks: These attacks involve stealing trained models from the system and using them for malicious purposes~\cite{74}.
\end{itemize}

Recent research explores various techniques to mitigate application-based attacks~\cite{10403965,10870125}. Researchers suggest using formal verification for smart contracts, implementing multi-signature controls, and deploying a hybrid PoW-PoS-BFT consensus mechanism to counteract logic exploits and 51\% attacks. In addition, AI-based real-time anomaly detectors, often integrated with digital twins or collaborative intrusion detection systems, are employed to filter the incoming sensor data and prevent the introduction of falsified information. Furthermore, models trained with adversarial robustness and continuous input verification are designed to resist crafted evasion attempts, whereas minor output perturbation along with rigorous query authentication are used to deter theft of proprietary models~\cite{10819009}.

\subsubsection{Quantum-based Attacks}
Quantum-based attacks in a 6G-ITS primarily target quantum-based technologies, such as Quantum Key Distribution (QKD), which might be incorporated into the system to ensure the security of communication and data transmission. These attacks can be categorized under data-based attacks, as they can compromise the system's security by breaking the cryptographic protection of classical key-based systems, leading to data breaches, safety hazards, and other issues~\cite{22}. Furthermore, quantum-based attacks can also affect the security of applications and devices within the 6G-ITS ecosystem, potentially compromising vehicle systems or introducing malicious software~\cite{40}. This highlights the cross-cutting nature of quantum-based threats and their ability to impact multiple layers of the 6G-ITS architecture.

Examples of quantum-based attacks include quantum key stealing, coherent attacks, Trojan-horse attacks, Measurement-Device-Independent (MDI) attacks, and quantum-safe key cloning attacks~\cite{22,139}. These attacks exploit the vulnerabilities of quantum-based technologies to intercept, manipulate, or clone cryptographic keys, bypassing the security measures meant to protect the 6G-ITS system.
Several techniques were proposed to mitigate the risk of quantum-based attacks across various layers of the 6G-ITS. These include implementing quantum-resistant cryptographic protocols, using quantum-safe key distribution methods, and implementing security measures such as access control and robust intrusion detection and response systems~\cite{40}. Additionally, incorporating quantum-based technologies such as quantum random number generators and quantum key distribution can help protect against these attacks~\cite{139}.
Acknowledging the cross-cutting nature of quantum-based attacks and their potential impact on multiple layers of the 6G-ITS architecture is crucial for developing a comprehensive security strategy. Implementing appropriate mitigation techniques across data security, application security, and device security can enhance the overall resilience of the 6G-ITS against evolving quantum-based threats.

\subsubsection{Privacy Invasion}
Privacy invasion poses a serious threat to 6G-ITS. The massive amounts of data collected, communicated, and analyzed in ITS can contain sensitive personal information. If exploited by attackers, this data could enable tracking of individuals' movements and routines, profiling of their behaviours and preferences, and other violations of privacy~\cite{5}. Several attributes make privacy attacks highly feasible in 6G-ITS systems. The ubiquity of sensors and cameras for traffic monitoring, vehicle tracking, enforcement purposes, fleet management and so on leads to pervasive surveillance~\cite{17}. The connectivity density in 6G networks allows this data to be easily aggregated and correlated~\cite{1}. Data mining and analytics techniques can deduce additional intelligence from raw data.

Key enablers of privacy attacks include visual sensors like cameras that can collect number plates or facial data, vehicular sensors gathering location and biometrics, and communication interfaces which can be exploited to track vehicles~\cite{17}. An example of such a privacy attack is a location inference attack demonstrated in our previous work on autonomous vehicle camera data, where we adopted a robust geo-localization technique to exploit subjects' location privacy in a distorted GAN-based camera dataset~\cite{111}. Side channel attacks that monitor radio signals can also reveal personal data. Even metadata derived from usage patterns and logs can erode privacy when aggregated and analyzed~\cite{113}.

Additionally, 6G capabilities will enable new big data business cases, such as monetization of vehicular data~\cite{112}. The huge amounts of data generated by connected and autonomous vehicles could be packaged and sold to third parties for purposes such as targeted advertising or insurance pricing. While this presents a revenue opportunity, it also poses major privacy risks if vehicle data is not properly anonymized and consented to. Travellers may not be comfortable having their driving patterns, vehicle health, and travel locations sold to private companies~\cite{114}. Strict opt-in consent and data protection controls would need to be implemented.

Defending against privacy breaches requires a combination of technical and policy interventions~\cite{112}. Anonymization and strong data encryption provide some safeguards, along with access restrictions. However, robust pseudonym management and identity protection schemes tailored for automotive scenarios are essential, given the scale and heterogeneity in 6G-ITS~\cite{71}. Differential privacy mechanisms can help prevent re-identification from statistical data releases~\cite{38}. Regulatory frameworks must evolve to address 6G-ITS specific challenges including the privacy implications of centimetre-level positioning accuracy, real-time behavioural analytics for traffic optimization, and the secure aggregation of massive sensor data streams across heterogeneous transportation networks.

\section{PRIVACY LANDSCAPE IN 6G-ITS}\label{sec:privacy}
Privacy protection in 6G-ITS faces unprecedented challenges as these systems collect vast amounts of personal data through continuous vehicle monitoring, behavioral tracking, and precise location sensing. This section explores privacy-preserving technologies such as federated learning and differential privacy, examines ethical considerations in data collection and monetization, and discusses frameworks for balancing system functionality with individual privacy rights.
\subsection{Privacy-Preserving Technologies and Mechanisms}
ITS uses advanced technologies such as sensors, communication networks, and data analytics to improve the efficiency and safety of transportation. These systems can generate large amounts of data about the movement of people and vehicles, which can be used for various purposes, such as traffic management, route optimization, and public safety~\cite{5}. Therefore, privacy in ITS can be classified into three categories: identity privacy, behaviour privacy and location privacy. Identity privacy pertains to the protection of user identification information, behaviour privacy involves safeguarding personal data generated by user actions, and location privacy pertains to the confidentiality of user location data~\cite{64, 38}. Privacy protection should be viewed as a crucial performance requirement and a fundamental component of wireless communication in the imagined era of 6G because 6G systems will have continuous connectivity up to around 1000 times larger than in 5G~\cite{72}. 

FL emerged as a promising solution. It is a decentralized machine learning paradigm enabling multiple devices to train a model collaboratively without sharing raw data with a central server~\cite{31}. FL offers several benefits, including preserving user privacy, reducing communication costs, and improving scalability. FL can help mitigate privacy risks by allowing data to be processed locally on user devices while enabling the system to learn from the collective intelligence of all users~\cite{100, 103}. FL can be used to train predictive models for traffic congestion and accident prediction, route optimization, and other ITS applications. Barbieri et al.~\cite{31} explored the potentials of the consensus-driven Federated Learning (C-FL) paradigm in V2X networks to provide communication-efficient distributed training services. C-FL typically performs well in dense networks with a large population of interconnected vehicles. However, The main challenges in FL design are device sampling, convergence and statistical heterogeneity, which can highly influence the quality of the trained models~\cite{102,103}. 
 
Future 6G wireless applications will likely incorporate Differential Privacy (DP), another emerging privacy-preserving technology~\cite{1, 5}. DP offers mathematically proven privacy protection against certain attacks, including inferencing, linkage, and reconstruction. Its properties, such as quantification of privacy loss, composition, and immunity to post-processing, make it attractive for enhancing privacy in analyzing personal information~\cite{64}. Lightweight privacy-preserving techniques like homomorphic encryption can also be used instead of traditional data encryption methods, providing the balance between maintaining the performance of high-accuracy services and the protection of user privacy~\cite{86}.

Trusted privacy preservation techniques require that communication participants have confidence in a third party to process their data~\cite{72}. A Central Authority (CA), which can link and invalidate user certificates used in encrypted communication, could be the independent legal authority. Examples include authentication, VPN/tunnel encryption, anonymization, and pseudonymization~\cite{90}. Several other privacy-enhancing technologies, such as Secure Multi-party Computation (SMC) and threshold secret sharing, offer prospective solutions to 6G-ITS privacy preservation~\cite{104}. One potential application of SMC in ITS is in the context of accident detection and response. SMC can enable real-time analysis of road conditions and detect accidents by securely aggregating data from different sources, such as vehicle sensors and camera cameras~\cite{106}. This can then inform the deployment of emergency services and traffic rerouting to minimize disruption. Moreover, in threshold secret sharing, a secret is divided into multiple shares, each distributed among different nodes in the network. The secret can only be reconstructed when a certain threshold of shares is collected~\cite{105,104}. This technique ensures that the secret remains secure even if some nodes are compromised.

In the 6G age, there is a greater risk that data collection and accessibility would undermine privacy protection and complicate regulatory issues. At the same time, edge intelligence aided by 6G changed the paradigm in how applications are developed and deployed. As a result, more sophisticated applications are being executed on resource-constrained mobile devices, increasing the risks of security attacks~\cite{86}. Consequently, safeguarding users' privacy on such devices has become a critical challenge. Therefore, it is imperative to incorporate lightweight and efficient privacy-preserving mechanisms. Maintaining a balance between the performance of high-accuracy services and user privacy protection is paramount. The realization of many intelligent applications requires access to sensitive user information, such as location and identity data. Hence, it is crucial to carefully consider data access rights and ownership and to establish effective mechanisms for supervision and regulation that ensure privacy protection.

\subsection{Ethical Considerations for 6G-ITS}
The deployment of 6G-ITS raises fundamental ethical concerns that extend beyond traditional cybersecurity and privacy considerations. User consent emerges as a critical challenge, as the unprecedented data collection capabilities of 6G networks will enable continuous monitoring of vehicle occupants, travel patterns, and behavioral characteristics with centimetre-level precision. Traditional opt-in consent mechanisms prove inadequate for dynamic 6G-ITS environments where data usage contexts evolve in real-time based on traffic conditions, emergency situations, and automated decision-making processes~\cite{SoK-Erfan10.1145/3709016.3737800}. The complexity of explaining AI-driven data processing to users in understandable terms further complicates informed consent, particularly when algorithmic decisions directly impact safety-critical transportation functions.

Algorithmic fairness and bias present additional ethical challenges as 6G-ITS systems increasingly rely on machine learning for traffic optimization, route planning, and resource allocation. These systems risk perpetuating or amplifying existing transportation inequalities if training data reflects historical biases in infrastructure development, service provision, or mobility patterns across different socioeconomic groups. The automated nature of 6G-ITS decisions makes bias detection and correction particularly challenging, especially when decisions occur within sub-millisecond timeframes that preclude human oversight~\cite{bossauer2020trust}. Real-world evidence already points to the sensitivity of these issues. Independent investigations, such as the Mozilla Foundation's 2023 report, highlight that modern connected vehicles can collect highly granular personal information, extending in some cases to intimate lifestyle inferences~\cite{caltrider2023cars}. Although such reports focus on current generation systems, the data collection capabilities in 6G-enabled ITS, combined with enhanced AI-driven analytics, will only amplify these risks if not governed by enforceable ethical frameworks.

Beyond these foundational concerns, emerging 6G-ITS architectures are likely to enable new economic models such as vehicle data marketplaces, where vehicle-generated data, ranging from traffic flows and sensor readings to in-cabin behavioral metrics, are monetized among stakeholders~\cite{Suo2020}. These models present potential incentives for data sharing and could fund infrastructure improvements or reduce vehicle costs; however, they simultaneously raise ethical and regulatory concerns about user consent, equitable distribution of value, and protection against exploitative practices~\cite{SoK-Erfan10.1145/3709016.3737800}. The inconsistency of global regulations further complicates these issues, as data monetization policies permissible in one jurisdiction may violate privacy rights in others, creating complex compliance challenges for global transportation systems. Operationalizing ethical frameworks for 6G-ITS requires integration of privacy-by-design at the architectural level, dynamic consent management that allows drivers to granularly approve or revoke specific data uses, and transparent revenue-sharing mechanisms that recognize drivers as active stakeholders rather than passive data sources. Additionally, algorithmic accountability measures must ensure that automated decisions can be audited, explained, and contested, particularly in safety-critical scenarios where system failures could have life-threatening consequences. Embedding these safeguards into the 6G-ITS design ensures that privacy protection moves beyond compliance checklists to a verifiable and enforceable practice that balances innovation with individual rights, even in heterogeneous regulatory and cultural landscapes.

\section{TRUST LANDSCAPE IN 6G-ITS}\label{sec:trust}
Trust is fundamental to the secure operation of 6G-ITS, where autonomous entities must reliably interact across heterogeneous, decentralized networks. This section explores the theoretical foundations of trust, examines governance frameworks and standardization efforts, and discusses key enabling technologies that shape trust management in next-generation intelligent transportation systems.
\subsection{Trust Fundamentals and Governance Framework}
Trust, while inherently abstract and challenging to define with precision, constitutes a fundamental component of interactions among distinct autonomous entities. In social environments, where individual behavior remains unpredictably variable, trust facilitates action by replacing the overwhelming intricacies of these contexts with predictive representations~\cite{142}. Thus, trust effectively reduces the infinite complexity of social scenarios to more manageable expectations about the probable conduct of others~\cite{125}.

The promise of 6G-powered ITS ecosystems is critical to engendering comprehensive trust spanning the technological, governance, and social spheres. The primary concerns revolve around establishing robust trust in the face of these novel integrations, ensuring reliable and secure communication in a dynamic and heterogeneous network environment~\cite{139}. The extensive connectivity and virtualization of 6G systems will require trusted mechanisms to manage service agreements and transactions on a massive scale. For example, blockchain solutions could facilitate decentralized billing and charging without intermediaries, as well as establish reliable service level agreements between virtual slices~\cite{138,137,131}. However, implementing such large-scale trust networks remains an open challenge~\cite{132}. In addition, ubiquitous sensors and data exchange in emerging intelligent transportation use cases raise critical privacy and security concerns. The sheer volume of spatial, visual, and contextual data from vehicle sensors enables unprecedented tracking and profiling~\cite{130}. If not protected properly, this data could enable harassment or stalking. Furthermore, manipulated sensor data injection could severely impact safety systems and trigger accidents. As such, stakeholders in the 6G ecosystem need stringent and resilient trust frameworks to preserve privacy, guarantee authenticity, and prevent misuse along the physical-digital boundary between vehicles, edge networks, and central clouds.

Decentralized networks such as VANET mobile networks require resilient trust management mechanisms that can function effectively despite vulnerabilities from within the network itself~\cite{125}. Traditional centralized methods of ensuring trust falter in such contexts. To enable robust decision-making at the node level, trust and reputation models must ingest direct evidence from local interactions and indirect inputs gathered through peer recommendations~\cite{126}. A trust calculation subsequently combines these multi-faceted evidence sources to determine a confidence score. Bootstrapping schemes handle the cold start problem by initializing new entities with neutral scores that get refined over time~\cite{125}. Various computational approaches power this evidence-gathering and trust calculation pipeline, including neural networks~\cite{142}, Bayesian~\cite{124}, entropy or probability-based methods~\cite{136}. The fusion of direct and indirect trust inputs lends hybrid resilience to decisions in decentralized environments. Operationalizing these schemes, however, remains an open research challenge.

As a result of these challenges, standardization bodies such as the Institute of Electrical and Electronics Engineers~(IEEE) and the European Telecommunications Standards Institute~(ETSI) formulated specifications around security and trust assurances in ITS. Notable among these efforts is ETSI TS 102 941~\cite{143}, which outlines a comprehensive trust framework spanning identities, cryptographic policies, and certification roots tailored to the automotive context. In its second iteration, this standard delineates trust relationships between ecosystem entities stipulates protocols for maintaining verified credentials necessary for secure communication and establishes collaborative trust anchor models, allowing multiple certificate authorities to interoperate within the same web of trust. The accompanying ETSI TS 102 940 guideline consolidates these specifications into a reference architecture that codifies best practices for trust establishment, maintenance, and cryptography for integrity and resilience. As ITS components become mainstream, rigorous adoption of such standards will be pivotal to managing trust across the complex partnerships between vehicles, infrastructure, networking components, and mobility service providers.
\noindent Table~\ref{tab:trustaspects} describes the trust aspects and their considerations regarding the 6G-ITS ecosystem.

\subsection{Key Enabling Technologies and Trust Implications}
While the technical details of 6G systems are still being researched and standardized, several key technology enablers have become clear. The emergence of the 6G network will provide breakthrough advances in speed, capacity, security, and intelligence compared to existing 5G infrastructure.  These capabilities have significant potential to foster trust in next-gen ITS.


\subsubsection{Artificial Intelligence and Machine Learning Technology}
Integrating AI/ML into the design and orchestration of next-generation 6G wireless networks is set to revolutionize intelligent transportation. The era of self-driving vehicles, fully connected infrastructure, and intelligent fleet coordination promises unprecedented change in how humans and goods travel - catalyzed by the unique capabilities of AI/ML~\cite{116}.

At its core, the explosive advancement of AI, especially deep neural networks, was driven by the availability of enormous data volumes, computational muscle, and ingenious algorithms modelled after human cognition~\cite{114,116}. 6G unlocks the mammoth connectivity bandwidth and ultra-low latency required to realize a fully AI-first future. Autonomous systems powered by reinforcement learning, decentralized edge intelligence ensuring lightning-fast response times, and generative models converting data into realistic environment simulations - all were hitherto constrained by 4G/5G limitations~\cite{80,107}.

The breadth of intelligent transportation use cases poised to be revolutionized by AI/ML in 6G networks is immense. Traffic flow optimization stands to become far more reliable through predictive modelling on rich datasets~\cite{11}. Entirely new forms of real-time vehicular coordination will be unlocked by AI-assisted V2X communication protocols~\cite{43, 119}. Self-driving technology could achieve responsiveness, rivalling human drivers via sensor fusion and computer vision~\cite{26}. Smart city infrastructure, from adaptive traffic signals to self-diagnosing components to automated parcel-sorting facilities, will grow more intelligent and efficient~\cite{21}.

A key shift is that 6G will enable more decentralized peer-to-peer communication, unlike current client-server models where the server is the presumed reliable component~\cite{47}. This requires participants to engage directly with one another through heterogeneous AI systems. Such multifaceted AI integration across smart transportation networks poses emerging challenges around interoperability, system-wide robustness, and auditability~\cite{51,61}. With automotive and transit functions becoming increasingly software-defined, rigorous validation of AI components and their interactions is essential to fostering trust at both the technical and social levels~\cite{132}. 

From a technical perspective, trust establishment begins with rigorous evaluation frameworks to quantify confidence in AI/ML components and multi-agent systems~\cite{135}. Quantitative trust metrics and benchmarks are needed, covering performance, robustness, security, explainability, and ethics~\cite{134}. Data veracity and integrity are other key trust factors - with decentralized aggregation, ensuring training data comes from reliable sources is imperative~\cite{135}. Ongoing trust management entails continuous auditing and adaptation as transportation environments and threats evolve~\cite{134}. 

Standards bodies have a role to play in devising interoperability and certification schemes for trustworthy AI agents to interoperate securely, particularly in safety-critical control functions~\cite{129,130}. Isolating mission-critical communications on verifiably secure network slices can mitigate risks~\cite{129}. Cybersecurity measures like encryption and access control remain essential to prevent spoofing, denial-of-service attacks, or hijacking. Several studies, e.g.,~\cite{131,122,123,137,121,132,136}, proposed methods to enhance trustworthiness in a 6G heterogeneous network. The authors of~\cite{131} proposed a scheme, ``6blokcs", where trust is managed and enhanced through a combination of blockchain technology, 6G sensors, and NFV. Blockchain ensures decentralized and transparent transactions, enhancing trust and security. The 6G sensors contribute to secure data aggregation, while NFV aids in efficient data processing and resource provisioning. Additionally, the system uses smart contracts for management operations, further ensuring security and trustworthiness in the network. 

The integration of AI, specifically Generative Adversarial Networks (GANs), into trust management for 6G networks is explored in~\cite{136}. It introduces a novel framework combining fuzzy logic and adversarial learning for intelligent trust management. The paper reviews existing AI-based trust management schemes, proposes a GAN-based trust decision-making model, and applies it to secure clustering for reliable and real-time communication. 

The requirement of incorporating trust into the 6G network's design is examined in~\cite{136}. It emphasizes the importance of trust in the merger of the digital and physical worlds, highlighting security risks and the need for robust security measures. The paper proposes a framework for embedding trust into the network, focusing on end-to-end connectivity and reputation-based trust management. It also compares current internet communication patterns with the proposed 6G framework, emphasizing the need for a paradigm shift in network trust and security approaches. 

Beyond technical factors, public trust relies on responsible and ethical development practices~\cite{132}. Humans tend to calibrate risks in AI failures, causing distrust~\cite{135}. Regulations addressing liability attribution, safety standards, and transparency requirements around the use of personal data are thus crucial. Social and legal frameworks such as the General Data Protection Regulation (GDPR) that protect public trust regarding digital privacy must be scaled to cover the additional use cases and business opportunities that 6G will birth. For example, data monetization platforms will need guidelines on who owns the generated data and who can sell and profit from processing such data. Also, insurance companies must determine who is liable if autonomous driving fails, resulting in a claim incident. Achieving these regulatory oversights and manufacturer liability provides accountability if mishaps occur. Therefore, AI/ML technologies promise to revolutionize 6G-ITS. Their success will largely depend on achieving a delicate balance between technical innovation and ethical, transparent practices. It is imperative to foster an environment where trust in delegating operations for AL/ML in 6G-ITS is as strong as the technology itself.

\subsubsection{Wireless Brain-computer Interaction}
The concept of wireless brain-computer communication is a topic of interest in the context of 6G-ITS. Wireless brain-computer communication involves using electroencephalography (EEG) signals to enable direct communication between the human brain and computer systems~\cite{30}. This technology can provide a more natural and intuitive way of interacting with technology, particularly in applications related to ITS. For example, wireless brain-computer communication could control various functions through brain signals, such as steering or braking. This emerging technology has the potential to revolutionize the way we interact with technology and our environment. With the development of 6G networks, there is an opportunity to explore the integration of wireless brain-computer communication in ITS applications~\cite{26}. Wireless brain-computer technology will enable brain-controlled vehicles (BCV) to help people with disabilities experience increased independence~\cite{95}. However, implementing wireless brain-computer communication in 6G-ITS applications also presents significant security, privacy, and ethical challenges. Given the sensitive and personal nature of the data involved, appropriate security measures are needed to protect against unauthorized access or manipulation. Additionally, ethical concerns must be considered, such as issues related to consent and the potential for unintended consequences~\cite{26,95}.

\subsubsection{Quantum Communication and Computing for 6G-ITS}
Quantum computing holds significant promise for bolstering trusted communications in 6G-ITS. By harnessing quantum mechanical phenomena like superposition and entanglement, quantum algorithms can enable cryptography to resist attacks from powerful future quantum computers~\cite{40}. Exploiting quantum parallelism and entanglement, quantum algorithms can break cryptography long considered unassailable, jeopardizing legacy security protocols~\cite{3,10}. Quantum key distribution offers a path to trusted communications that are resistant even to quantum attacks~\cite{140}. However, seamless integration with classical networks and devices remains an open challenge~\cite{139}.

On the flip side, quantum computing could bolster security and trust mechanisms with new capabilities. Quantum machine learning shows potential for anomaly detection in complex real-time data, identifying early indicators of malfunction or intrusion~\cite{117}. Entangled quantum sensor networks may achieve orders of magnitude enhancement in sensitivity for integrity checks on critical infrastructure~\cite{141}. And post-quantum cryptographic primitives theoretically outside the reach of quantum brute forcing are rapidly maturing~\cite{40,140}. Navigating this quantum computing nexus requires a multilayered approach balancing legacy compatibility, future-proofing, and pragmatic transition. Hybrid asymmetric schemes mixing quantum-safe and conventional cryptography provide a migration path as post-quantum standards co-evolve with the technology. Isolating security-critical applications like V2X coordination onto verifiably secure network slices can contain risks~\cite{140}. Blockchain-based consensus offers decentralized trust roots that are less dependent on computational hardness assumptions~\cite{141}.

The quantum paradigm necessitates upgrading computational trust protections for 6G transportation foundations. But judiciously leveraged, quantum techniques can also significantly improve resilience, confidentiality, and reliability - promoting user acceptance. Collaborative efforts between automotive, telecom and quantum stakeholders provide the ideal environment for co-designing trusted mobility networks. The duality of quantum computing's impact means that while it can greatly improve the efficiency and security of 6G-ITS, it also necessitates a proactive approach to updating and fortifying current cryptographic standards to maintain network integrity and trust.

\subsubsection{Advance Wireless Communication Technonlogies}
The development of advance wireless communication technologies like terahertz~\cite{42}, light fidelity (Li-Fi)~\cite{44}, cell-free massive multiple input multiple outputs (MIMO)~\cite{11}, and intelligent reflecting surfaces (IRS)~\cite{23} will enable ultra-fast, secure, and reliable connectivity for future 6G-ITS. 

Terahertz band communication is a promising wireless technology for 6G that can enable the high data rates and wide bandwidths needed for real-time coordination between vehicles, infrastructure, and devices in ITS~\cite{42}. By operating at frequencies between the mmWave and infrared bands, terahertz provides far greater spectrum resources than existing wireless technology. This vast bandwidth can satisfy the capacity demands of future mobility ecosystems, potentially reaching transmission speeds up to terabits per second~\cite{42,11}. The tiny wavelengths of terahertz signals also permit the integration of thousands of antenna elements into compact base station arrays~\cite{11}. This massive MIMO capability allows the forming of highly directional beams to service many simultaneous users while minimizing co-channel interference reliably~\cite{26}.

From a trust and security perspective, the focused energy and limited penetration of terahertz beams inherently resist eavesdropping, enhancing privacy~\cite{24}. This prevention of unauthorized access aligns with integrity goals for safety-critical vehicle-to-everything communication. However, designing efficient components like antennas and transceivers at such high frequencies remains challenging~\cite{26}. Further research is needed to make terahertz systems economically viable at scale. And seamless integration with networks at lower frequencies will be crucial for ubiquitous coverage across transportation infrastructure. Suppose the complexity and power efficiency barriers can be overcome. In that case, terahertz communication promises to provide the secure ultra-high capacity connectivity required to unlock 6G's potential for next-generation intelligent mobility.

Visible light communication (VLC) is an emerging wireless technology that uses LED light sources for high-speed wireless data transmission~\cite{44}. By modulating visible light signals, VLC can offer fibre-optic-like bandwidth without requiring the installation of physical cables~\cite{26}. This makes it highly promising for vehicle-to-vehicle and vehicle-to-infrastructure communication within intelligent transportation ecosystems.

Since VLC signals do not penetrate opaque objects, connections are inherently confined to intended recipients within direct line-of-sight~\cite{44}. This localization enhances security and privacy compared to RF links~\cite{68}. Interference is also avoided by spatial reuse of frequencies~\cite{44}. With thousands of LEDs integrated into vehicles and smart infrastructure, VLC could enable terabit aggregate data rates for real-time coordination and content sharing as required by future autonomous mobility services. However, some challenges remain in efficiently harnessing VLC for transportation. Commercial LEDs have narrow modulation bandwidths, requiring enhancement~\cite{44}. Robust multi-input multi-output techniques must be developed to overcome limited diversity vulnerabilities~\cite{68}. Seamless integration with RF networks is needed where line-of-sight is obstructed. VLC promises to provide the speed, security, and reliability needed to establish trusted links between myriad intelligent devices across 6G transportation ecosystems.

Massive MIMO beamforming is an emerging wireless technique that utilizes large antenna arrays for highly directional signal transmission and reception~\cite{44,24}. By steering focused energy beams, massive MIMO enables spatial reuse where multiple users can simultaneously utilize the same frequencies without interference~\cite{49}. This significantly enhances spectral efficiency and capacity compared to omnidirectional broadcasts.

For 6G intelligent transportation, massive MIMO beamforming offers important trust and security advantages. Focused directional beams provide reliable, ultra-high-speed links between vehicles, infrastructure, and devices by confining signals only where needed. This resists eavesdropping or data interception, enhancing privacy~\cite{49,43}. The high gain also compensates for path loss at higher frequencies like mmWave and terahertz, which are envisioned for 6G~\cite{43}. Additionally, interference is minimized as directional transmissions avoid overlapping with other links~\cite{29}. While promising, real-world mobility environments pose challenges. Maintaining precise beam alignment requires tracking mobile nodes and adapting beams dynamically~\cite{23}. Blockage from buildings or vehicles can disrupt links~\cite{23}. Scaling up low-cost antenna arrays with reliable phase synchronization is non-trivial. And seamless integration with omnidirectional networks is necessary to prevent coverage gaps~\cite{49}. If these hurdles can be overcome, massive MIMO beamforming is poised to deliver the speed, security, and scalability needed to realize the trusted connectivity potential of 6G transportation.

Distributed MIMO (dMIMO) extends traditional MIMO systems by decentralizing antenna arrays across multiple access points (APs)~\cite{155}. This architecture enhances signal reliability, coverage, and capacity, making it suitable for 6G-ITS~\cite{153}. dMIMO's cooperative nature enables dynamic adaptation to channel variations, ensuring efficient resource use and mitigation of interference. dMIMO can support high-density vehicular networks by providing more uniform coverage and reducing the impact of signal blockages. By strategically placing antenna elements throughout the transportation infrastructure, dMIMO can ensure reliable vehicle connectivity, even in challenging environments like urban canyons or tunnels~\cite{152,154}. The distributed nature of dMIMO also allows for more efficient use of power, as antenna elements can be dynamically activated or deactivated based on traffic demands and vehicle locations~\cite{151}.

ISAC is a crucial enabler in 6G networks, integrating sensing with communication infrastructure for efficient spectrum use and enhanced data accuracy in ITS operations~\cite{146}. This integration reduces the attack surface by having fewer points of potential compromise compared to separate systems~\cite{144}. ISAC employs advanced encryption and secure signal processing to protect data integrity and prevent eavesdropping, though it must also guard against spoofing attacks that could mislead sensing functions~\cite{147}. The environmental data collected by ISAC can potentially be exploited for unauthorized tracking. Ensuring privacy requires robust data anonymization and strict access controls. Designing ISAC systems with privacy-by-design principles, such as minimal data collection and end-to-end encryption, helps mitigate privacy concerns~\cite{149,150}. Trust in ISAC is built on system reliability and data accuracy. Bidirectional authentication ensures that devices and infrastructure verify each other’s identities, preventing attacks from fake base stations~\cite{150,151}. Additionally, machine learning algorithms can detect anomalies and security threats in real-time, further enhancing trust~\cite{144}.

Intelligent reflecting surfaces (IRS) are an emerging technology comprised of software-controlled metamaterials that can dynamically alter how impinging radio waves are reflected and directed~\cite{}. By adjusting the phase shifts applied by a dense array of IRS elements, wireless signals can be precisely steered to intended recipients while minimizing undesired scattering. This allows wireless environments to be reconfigured and optimized in real-time based on changing conditions~\cite{82}.

In 6G-ITS, IRS offers promising opportunities to isolate trusted device groups selectively, reinforce signal strength only where needed, and null interference or eavesdroppers~\cite{23}. IRS could also enable rapid beam retargeting for reliable vehicle-to-infrastructure links during high-speed mobility~\cite{82}. Integration of IRS with AI/ML and massive MIMO techniques can further augment adaptivity, tuning propagation intelligently. However, seamless coordination across large dynamic IRS networks poses challenges, including channel modelling, optimal phase configuration, mobility management, and cost-effective fabrication~\cite{23,82}. The software-defined nature of the IRS also introduces potential vulnerabilities if not properly secured. Failures or misconfigurations could disrupt critical links~\cite{23}. Moreover, the complexity of unpredictable signal interactions in complex transportation environments may limit real-world deployments~\cite{82}. Several papers~\cite{156,157,158} highlighted the security threats and vulnerabilities associated with IRS-aided networks. These threats include jamming attacks using adversarial IRSs, pilot contamination attacks (PCA) exploiting IRS to enhance eavesdropping capabilities, and metasurface manipulation attacks (MSMA) manipulating metasurface behaviour for malicious purposes. The implications of these threats on the IRS are substantial, as they expose the vulnerability of IRS-aided wireless systems to various forms of attacks that can compromise the security, privacy, and reliability of communications, highlighting the need for developing robust security measures and countermeasures specifically designed for IRS-aided networks. While still in a developmental phase, IRS represents a revolutionary paradigm shift that could give 6G communications unprecedented control over the wireless medium.

The advanced wireless communication technologies discussed above, such as terahertz, VLC, massive MIMO beamforming, dMIMO, ISAC, and IRS, offer promising solutions for ultra-fast, secure, and reliable connectivity in 6G-ITS. However, as these technologies evolve and become more complex, ensuring the security, privacy, and trust of the transmitted information becomes increasingly critical. Physical layer security (PLS) emerges as a crucial first line of defence, exploiting the inherent randomness of wireless channels to secure communications without relying solely on traditional cryptographic methods~\cite{161}. PLS schemes were proposed for various 5G and beyond technologies, aiming to enhance the security performance of heterogeneous networks, device-2-device (D2D) communications, IoT, and other aspects of 6G-ITS~\cite{162}. The integration of PLS with advanced techniques such as non-orthogonal multiple access (NOMA), IRS, and machine learning showed great potential to achieve better security performance~\cite{159,162}. One promising approach to improving spectrum and energy efficiency in 6G networks is the integration of NOMA and Reconfigurable Intelligent Surface (RIS) techniques. Studies investigated the PLS for RIS-aided NOMA 6G networks, considering both internal and external eavesdropping scenarios. Researchers proposed joint beamforming and power allocation schemes to improve system PLS when dealing with untrusted near-end users attempting to intercept far-end user information. In both internal and external eavesdropping scenarios, noise beamforming and optimal power allocation schemes were introduced to enhance the system's physical security, even without channel state information (CSI) for the eavesdroppers~\cite{160}. However, the practical implementation of PLS in 6G-ITS faces challenges, such as managing security across diverse devices, efficient resource allocation, and adapting to dynamic environments~\cite{159,163}. Future research should focus on addressing these challenges and developing comprehensive security frameworks that integrate PLS with higher-layer security measures to ensure the trustworthiness of 6G-enabled intelligent transportation systems.

\subsubsection{Mobile Edge Computing (MEC)}
The proliferation of intelligent mobility applications relying on immense volumes of latency-sensitive vehicular data warrants deeper edge computing integration in 6G ecosystems through multi-access edge computing (MEC) platforms~\cite{137}. By embedding compute, storage and analytics closer to autonomous vehicles, drones and transportation infrastructure, sub-millisecond critical decision-making can be facilitated while smoothing core cloud burdens~\cite{46}. MEC’s localized intelligence also aids predictive traffic optimization, vehicle coordination, and dynamic network resource allocations to serve native contextual demands better~\cite{88}. However, the radically enlarged edge attack surface risks magnifying vulnerabilities~\cite{103}. Ensuring reliable hardware roots-of-trust, compartmentalization integrity between slices sharing resources, and securing inter-edge mobility handoffs grows pivotal. With autonomous vehicles bearing lives, prudent threat modelling spanning communication and computing techniques that smartly fuse edge autonomy with cloud supervision minimizes risks.

\subsubsection{Blockchain and Distributed Ledger Technology (DLT)}
Blockchain and distributed ledger technologies offer promising trust mechanisms for 6G transportation ecosystems by establishing decentralized consensus and immutable transaction records between entities~\cite{138}. DLTs provide transparency and prevent data tampering or forgery, aligned with security and integrity goals~\cite{137}. Blockchain will facilitate the transition from centralized client-server architectures to trusted peer-to-peer networks for intelligent transportation. As described in~\cite{138,121}, blockchain and distributed ledger technologies (DLT) can enable the next generation of distributed sensing and coordination for advanced mobility services based on decentralized trust mechanisms rather than centralized intermediaries.

To fully realize the potential trust benefits, the demanding connectivity requirements of blockchain-based transportation applications will necessitate a synergistic combination of capabilities enabled by 6G networks~\cite{131,121}. Ultra-reliable, low-latency communications will provide the real-time data exchange and consensus needed to foster dynamic trust between vehicles, infrastructure, and other endpoints~\cite{121}. Massive machine-type communications will deliver the scalability to support potentially billions of transported entities and infrastructure endpoints within these trust frameworks~\cite{120}. There are several opportunities for implementing blockchain and DLT in 6G-ITS:
\begin{itemize}
    \item Decentralized identity management for vehicles, drivers, and infrastructure components enables authenticated and authorized access to transportation networks. DLTs allow interoperable yet privacy-preserving identification.
    \item Secure over-the-air software updates, where the integrity of firmware upgrades and vehicle vulnerability patches can be verified through blockchain-based provenance tracing and code attestations.
    \item Supply chain transparency and efficiency improvements via shared ledgers tracking location, condition and freight handling in real-time across multi-party logistics flows.
    \item Insurance claim processing can leverage automated execution of coverage rules and payment settlements via smart contracts in case of accidents, delays and vehicle failures.
    \item Electric vehicle charging and billing mediated through cryptocurrency payments \cite{BahramiLIoTning2025} without a central clearing house using Vehicle-to-Grid integration platforms.
    \item 5G/6G network slice brokers governed decentralized by blockchain for on-demand quality-of-service guarantees to latency-sensitive or high throughput applications.
    \item Mesh networks and spectrum sharing orchestrated securely between transportation stakeholders, leveraging distributed consensus and incentives for cooperative behaviours.
    \item Community sensor data exchange platforms to share traffic, weather or pollution alerts in common ledgers while preserving data ownership rights.
    \item vehicular data monetization \cite{SoK-Erfan10.1145/3709016.3737800} whereby connected vehicle owners can take ownership of their data and trade it in a peer-to-peer manner.
\end{itemize}
As 6G systems scale exponentially in complexity, spanning vehicles, infrastructure, and supply chains - blockchain and DLTs can crucially offer transparency, auditability, automation and coordination across such intricately interconnected ecosystems in a trusted manner.
\noindent Table \ref{tab:6G} enumerates the opportunities, privacy and security challenges of key 6G enabling technologies and their application in the ITS environment.

\definecolor{verylightgray}{gray}{0.95}
{
\renewcommand{\arraystretch}{1.6}

\begin{table*}[!ht]
    \centering
    \caption{Opportunities and challenges of 6G-enabling technologies for ITS.}
    \label{tab:6G}
    \resizebox{\linewidth}{!}{
    \begin{tabular}{|>{\centering\arraybackslash}m{2.4cm}|>{\arraybackslash}m{3.8cm}|>
    {\arraybackslash}m{2.9cm}|>
    {\arraybackslash}m{3cm}|>
    {\arraybackslash}m{3.5cm}|>
    {\arraybackslash}m{3.5cm}|}
    \rowcolor{verylightgray}
    \hline 
    \textbf{6G-enabling technology} & \multicolumn{1}{c|}{\textbf{Opportunities for ITS}} & \multicolumn{1}{c|}{\textbf{Arising privacy concern}} & \multicolumn{1}{c|}{\textbf{Security challenge}} &  \multicolumn{1}{c|}{\textbf{Use cases}} & \multicolumn{1}{c|}{\textbf{Trust issue}} \\ \hline

    Artificial Intelligence (AI)& Enhanced predictive intelligence and real-time decision making for autonomous systems, leveraging 6G's ultra-low latency and massive connectivity & Data collection, Data Anonymization, Data sharing & Increased complexity of AI models in 6G environments, Requiring more sophisticated security measures & Intelligent traffic management, Autonomous driving, Route and planning optimization, AI-driven network optimization and self-healing in 6G ITS infrastructure & Ensuring transparency and accountability of AI systems in ultra-dense, high-speed 6G networks
    \\ \hline
    
    Wireless Brain-Computer Interactions & Driver Assistance, Accessibility, Cognitive workload management & Data storage and retention, Unauthorized access, Data disclosure & Authentication, Authorization, Integrity & Remote driving, Advanced Driver Assistance Systems (ADAS), Personalized in-vehicle entertainment & Trust in direct brain-computer interfaces, Privacy risks from neural data collection and use. Reliability and accuracy of brain-computer interface interpretations.
    \\ \hline
         
    Quantum Computing & Optimization, Quantum sensors, Cybersecurity   & Privacy-preserving computation, Data ownership & Malware attacks, Access control & Routing and navigation, V2X &  Trustworthy integration with classical networks \\ \hline
         
         Terahertz Communication (Thz) & High-speed communication, Improved sensing capabilities, Efficient spectrum utilization & Data retention, Unintentional radiation & Authentication, Interference, Encryption & Advanced driver assistance systems (ADAS), V2V, V2I & Reliability and consistency of Thz communication in diverse environments and maintaining privacy against potential eavesdropping due to its high-frequency signals.\\ \hline
         
         
         Mobile Edge Computing (MEC) & Ultra-low latency edge processing and analytics, enabled by 6G's increased bandwidth and reduced latency & User profiling, Data breaches & Authentication, Access control and securing the vastly increased number of edge nodes in 6G networks & Real-time, AI-powered traffic management utilizing 6G's massive machine-type communications & Maintaining data integrity and privacy across a highly distributed 6G edge computing environment \\ \hline
         
         Intelligent Reflective Surface (IRS) & Improved signal quality, Energy efficiency, Enhanced communication range & Location tracking, Data collection & authentication, confidentiality & V2I, Smart intersection management, Autonomous vehicle navigation & Secure coordination across large IRS networks \\ \hline
         
         Blockchain and Distributed Ledger Technology (DLT) & Secure and reliable data sharing, Smart contracts for autonomous vehicles, Data security and privacy & Pseudonymity, Linkability & 51\%attacks, Lack of regulation & Traffic management, Vehicle identity and ownership, Decentralized ride-sharing & Consensus integrity, establishing trust in decentralized systems, reliability in SLA automation \\
         \hline
        Network Function Virtualization (NFV) & Flexibility, Scalability, Agility, Resiliency & Virtual appliance vulnerabilities & Expanded attack surface, Misconfiguration & On-demand network scaling, Hardware abstraction & Securing virtualized functions and pipelines \\ \hline
    \end{tabular}
    }
\end{table*}
}



\subsubsection{Network Function Virtualization (NFV)}
Regarding infrastructure, NFV is expected to play a significant role in 6G-ITS by enabling the virtualization of network functions traditionally carried out by proprietary physical devices~\cite{108}. The flexibility and agility offered by NFV can provide benefits such as efficient computing resource utilization, faster deployments and updates, reduced costs, dynamic scaling to meet fluctuating demands, and enhanced resiliency through virtualized redundancy and failover capabilities~\cite{108,109}. The scalability and flexibility inherent in NFV provide a robust framework for ITS to adapt to varying traffic conditions in real-time~\cite{107}. This dynamic allocation of resources enables more efficient use of network capabilities, which is a crucial factor in modern, fast-paced, heterogeneous transportation systems. However, research showed that the vastly expanded attack surface from software pipelines, compatibility challenges posed by multi-vendor virtual network functions (VNFs), and resource isolation needs pose complex trust management challenges~\cite{110,125}.

With network slicing, isolated virtual slices can be devoted to specific use cases or tenants. This permits customized trust settings and service levels by logical partition rather than a one-size-fits-all network~\cite{126}. Slicing also aids scalability for massive device density. However, the programmability of NFV presents attack vectors like configuration tampering. Hardening virtualized infrastructure against threats is crucial. Distributed ledger-based platforms show promise in establishing immutable configurations and managing trust in software-defined ecosystems~\cite{130}.

\definecolor{verylightgray}{gray}{0.95}

{
\renewcommand{\arraystretch}{2}

\begin{table*}
    \centering
    \caption{Trust aspects and considerations in 6G-ITS}
    \label{tab:trustaspects}
    \resizebox{\linewidth}{!}{
    \begin{tabular}{|>{\centering\arraybackslash}m{1.5cm}|>{\centering\arraybackslash}m{4cm}|>
    {\centering\arraybackslash}m{4.8cm}|>{\centering\arraybackslash}m{5.27cm}|>{\centering\arraybackslash}m{4.28cm}|}
\rowcolor{verylightgray}
    \hline 
\textbf{Aspect} & \textbf{Key Technologies} & \textbf{Trust Challenges} & \textbf{Trust Solutions} & \textbf{Application Scenarios} \\
\hline 
Technological trust & 
    \begin{itemize}[leftmargin=*, label=\textbullet, before=\vspace{3pt}]
        \item AI/ML
        \item Blockchain / DLT
        \item Quantum Computing
        \item Advanced Wireless (THz, VLC, MIMO, IRS)
    \end{itemize} & 
    \begin{itemize}[leftmargin=*, label=\textbullet, before=\vspace{3pt}]
        \item Interoperability
        \item System-wide robustness
        \item AI bias and explainability
        \item Quantum-safe cryptography integration
    \end{itemize} & 
    
    \begin{itemize}[leftmargin=*, label=\textbullet, before=\vspace{3pt}]
        \item Rigorous evaluation frameworks
        \item Quantum-resistant algorithms
        \item Federated learning
        \item Physical layer security
    \end{itemize} & 
    
   \begin{itemize}[leftmargin=*, label=\textbullet, before=\vspace{3pt}]
        \item Autonomous vehicle coordination
        \item Smart traffic management
        \item Predictive maintenance
    \end{itemize} \\
    
\hline
Network trust & 
    \begin{itemize}[leftmargin=*, label=\textbullet, before=\vspace{3pt}]
        \item NFV    {\hspace{1.3cm}\textbullet \hspace{0.2cm}VLC}
        \item MEC    {\hspace{1.25cm}\textbullet \hspace{0.2cm}ISAC}
        \item dMIMO  {\hspace{0.92cm}\textbullet \hspace{0.2cm}IRS}
    \end{itemize} & 
    \begin{itemize}[leftmargin=*, label=\textbullet, before=\vspace{3pt}]
        \item Expanded attack surface
        \item Resource isolation
        \item Edge node security
        \item Frequency interference and jamming
    \end{itemize} & 
    \begin{itemize}[leftmargin=*, label=\textbullet, before=\vspace{3pt}]
        \item Network slice isolation
        \item Distributed ledger for configurations
        \item Zero-trust architecture
    \end{itemize} & 
    \begin{itemize}[leftmargin=*, label=\textbullet, before=\vspace{3pt}]
        \item V2X communication
        \item Remote surgery
        \item Augmented reality navigation
    \end{itemize} \\
\hline
Data trust & 
    \begin{itemize}[leftmargin=*, label=\textbullet, before=\vspace{3pt}]
        \item Blockchain
        \item Federated Learning
        \item Homomorphic Encryption
    \end{itemize} & 
    \begin{itemize}[leftmargin=*, label=\textbullet, before=\vspace{3pt}]
        \item Data privacy
        \item Data integrity
        \item Unauthorized access
    \end{itemize} & 
    \begin{itemize}[leftmargin=*, label=\textbullet, before=\vspace{3pt}]
        \item Differential privacy
        \item Secure multi-party computation
        \item Blockchain for data provenance
    \end{itemize} & 
    \begin{itemize}[leftmargin=*, label=\textbullet, before=\vspace{3pt}]
        \item Vehicular data monetization
        \item Personalized traffic services
        \item Collaborative sensing
    \end{itemize} \\
\hline
Governance trust & 
    \begin{itemize}[leftmargin=*, label=\textbullet, before=\vspace{3pt}]
        \item Smart Contracts
        \item Decentralized Identity
    \end{itemize} & 
    \begin{itemize}[leftmargin=*, label=\textbullet, before=\vspace{3pt}]
        \item Regulatory compliance
        \item Cross-border interoperability
        \item Liability attribution
    \end{itemize} & 
    \begin{itemize}[leftmargin=*, label=\textbullet, before=\vspace{3pt}]
        \item Standardization (e.g., ETSI TS 102 941)
        \item Regulatory frameworks (e.g., GDPR for 6G)
        \item Automated compliance checking
    \end{itemize} & 
    \begin{itemize}[leftmargin=*, label=\textbullet, before=\vspace{3pt}]
        \item Automated insurance claims
        \item Dynamic road pricing
        \item Cross-border vehicle authentication
    \end{itemize} \\
\hline
{Social trust} & 
    \begin{itemize}[leftmargin=*, label=\textbullet, before=\vspace{3pt}]
        \item Explainable AI
        \item Privacy-Enhancing Technologies
        \item Human-Machine Interfaces
    \end{itemize} & 
    \begin{itemize}[leftmargin=*, label=\textbullet, before=\vspace{3pt}]
        \item Public perception
        \item Ethical considerations
        \item User acceptance
    \end{itemize} & 
    \begin{itemize}[leftmargin=*, label=\textbullet, before=\vspace{3pt}]
        \item Transparent AI decision-making
        \item User-centric design
        \item Ethical guidelines and frameworks
    \end{itemize} & 
    \begin{itemize}[leftmargin=*, label=\textbullet, before=\vspace{3pt}]
        \item Trust-based ride-sharing
        \item Ethical routing choices
        \item Community-based traffic reporting
    \end{itemize} \\
\hline
    \end{tabular}
    }
\end{table*}
}

The security of ITS enabled with 6G and enabled with 5G is similar in that both systems must protect the communication and data being transmitted, processed and stored against various threats and attacks. However, there are several key differences between the security of 6G-ITS and 5G-enabled ITS.

One of the main differences between 6G-ITS and 5G-enabled ITS is the level of security required to protect high-speed and low-latency 6G communication~\cite{15,75}. The increased data transmission speeds and low latencies of 6G would require more advanced security measures to protect against sophisticated cyberattacks and ensure the integrity and authenticity of the data. The application of new technologies is also a significant difference. mmWave and THz communications, which are relatively new technologies that are expected to be used in 6G, would present additional security concerns, such as defending communication channels and devices from jamming and spoof attacks~\cite{27,39,76}.

Incorporating quantum-based technology is another distinction, necessitating new and sophisticated security methods like Quantum Key Distribution (QKD) to secure communication channels and quantum-resistant algorithms to thwart quantum-based cyberattacks. 5G technology successfully shaped the development of the next generation of ITS and mapped the demands required for integration and adoption with the cellular network through softwarization; however, there is still concern for security around the 5G architecture~\cite{77}. Another significant obstacle to the effectiveness and precision of security controls and monitoring solutions is the high degree of heterogeneity in the 5G-ITS network~\cite{8}. 5G-ITS must support various devices and a lot of network traffic. A network of this size can increase the attack surfaces and allow threats to travel to the main areas of the connectivity; this threat is also present in 6G-ITS, as 6G will support devices more loosely connected in a short range than 5G~\cite{5}. Consequently, it raises important questions about how to build reliable connections between devices and networks.

Currently, 5G networks are adequate for existing technologies~\cite{80}. However, limitations in speed, manual configuration, and network optimization make it unsuitable for supporting future applications and scaling up to accommodate many complex dynamic wireless networks~\cite{35,51}. For example, emergency services in the ITS ecosystem need live feeds in almost real-time to attend to a situation remotely using unmanned vehicles (UVs). This usually requires information transmission at a latency speed below 1ms; however, 5G offers only a 5ms latency~\cite{60}. Furthermore, as ITS services and applications become more integrated, the number of loosely connected nodes that must communicate as a unit is on the rise; therefore, network congestion could be a risk if 5G is unable to link such a number of devices, resulting in further delay in information transmission~\cite{46}. To avoid serious CAV connectivity failures, the 5G network topologies will be significantly based on the applications, quality monitoring, and security of the service providers~\cite{35}. Security must be carefully considered with 5G relying on identical mobile network networks and utilizing virtualization and multi-tenancy capabilities. The multiple levels of security and privacy concerns in 5G networks are as follows: The vulnerability between the gNB radio node and the Mobility Management Entity (MME) due to modified nodes exists in backhaul lines~\cite{76}, core networks are more dynamic and susceptible to attacks due to SDN, NFV and cloud approaches, and edge networks are more vulnerable due to the heterogeneity of nodes and transitions between different access technologies~\cite{46,78}.

\subsubsection{Open Radio Access Network}
Open Radio Access Network (O-RAN) represents a transformative enabler for 6G-enabled ITS through its disaggregated architecture, open interfaces, and intelligent automation capabilities that offer unprecedented opportunities for transportation systems~\cite{10024837,9839628}. The flexibility of the O-RAN architecture enables multi-vendor environments with cost-effective deployments. At the same time, intelligent Radio Intelligent Controllers (RICs) provide real-time network optimisation and resource management critical for ultra-reliable low-latency communications in autonomous vehicle coordination and V2X applications~\cite{10937198}.  Near-RT RIC facilitates millisecond-level decision-making through xApps that can implement traffic-aware handover algorithms, predictive resource allocation, and intelligent interference management essential for 6G ITS scenarios~\cite{10829659}. Furthermore, O-RAN's open interfaces promote vendor interoperability and innovation, enabling rapid deployment of AI/ML-driven network functions for traffic management, predictive maintenance, and adaptive quality of service provisioning in transportation networks~\cite{10969847}. However, these opportunities come with substantial security challenges, as comprehensive threat analysis reveals 68 distinct threats with approximately 76\% classified as high-risk~\cite{10606000, oran2022threatmodel}. Critical vulnerabilities include malicious xApp deployment that can compromise sensitive user data, launch adversarial attacks against AI/ML models, or propagate infections across network functions~\cite{oran2022threatmodel}. The lack of control plane encryption in the open fronthaul interface allows for man-in-the-middle attacks and impersonation of network elements. Additionally, weak authentication mechanisms across O-RAN APIs make systems vulnerable to credential theft and service disruptions~\cite{10606000}. These security challenges are particularly concerning for 6G ITS, where O-RAN vulnerabilities could enable attackers to manipulate traffic management systems, compromise vehicle communications, or disrupt emergency response coordination~\cite{10807044}.

To effectively address these security challenges, zero trust architecture (ZTA) principles should be adopted, assuming the presence of continuous threats and requiring explicit authentication for all entities in the network, with real-time monitoring and dynamic risk assessment~\cite{10807044, 10024837}. Essential security measures include the deployment of a robust PKI for mutual authentication between O-RAN components, the implementation of Transport Layer Security (TLS) encryption on all interfaces, and the establishment of rigorous verification frameworks for xApp, utilizing Role-Based Access Control (RBAC) and Sandboxing techniques~\cite{10330565, fi17060233}. Advanced security technologies, such as Post-Quantum Cryptography (PQC) algorithms, provide quantum-safe protection for long-term 6G ITS deployments. Meanwhile, blockchain-enabled, decentralized identity management ensures tamper-proof authentication and authorization~\cite{10807044, 139}. Physical layer security mechanisms can provide additional protection against eavesdropping and jamming attacks specific to transportation environments~\cite{10937198}. Given the safety-critical nature of ITS applications, O-RAN security implementations must maintain ultra-low latency and high reliability requirements through optimized security protocols and hardware-accelerated cryptographic operations~\cite{10807044}. Future research should focus on developing AI-driven threat detection capabilities that leverage transformer-based models for real-time anomaly detection, formal verification methods for xApp security to ensure mathematical guarantees of application behaviour, and standardized security frameworks specifically designed for O-RAN deployments in transportation contexts to ensure consistent protection across multi-vendor environments~\cite{10829659, 10872859}.

\noindent Table~\ref{tab:5g6gcom} presents a systematic comparison between 5G-enabled ITS and 6G, illustrating how technological advances enable enhanced security capabilities. Each feature comparison is accompanied by its cybersecurity implications and practical application scenarios. For example, while the increase in data transmission speeds from 20 Gbps to 1 Tbps primarily appears as a performance metric, it necessitates new security protocols to ensure data integrity and timely encryption for applications like remote driving. Similarly, the reduction in latency to sub-1 ms creates new security considerations around timing attacks that must be addressed for safe autonomous vehicle coordination. This interconnected analysis demonstrates how 6G's technological foundations both enable and require advanced security measures beyond those in 5G-ITS.

\section{5G-ITS VS 6G-ITS SECURITY}\label{sec:5Gvs6G}
The progression from 5G to 6G networks in ITS represents a significant evolution in security requirements and capabilities. Table~\ref{tab:5g6gcom} provides a comprehensive comparison of the security implications of 5G and 6G-ITS and application scenarios. This systematic analysis shows how each technological advancement in 6G not only enhances network capabilities but also introduces new security considerations that must be addressed. Building on this comparative foundation, the following section examines the specific challenges that emerge in 6G-ITS environments, particularly focusing on the fundamental issues of authentication and trust establishment in next-generation transportation networks.

\definecolor{verylightgray}{gray}{0.95}
{
\renewcommand{\arraystretch}{1.6}
\begin{table*}[htbp]
\centering
\caption{Comparison of 5G and 6G-ITS security implications and application scenarios.}
\label{tab:5g6gcom}
\resizebox{\linewidth}{!}{
\begin{tabular}{|>{\centering\arraybackslash}m{2.2cm}|>{\centering\arraybackslash}m{2.2cm}|>
{\centering\arraybackslash}m{2.6cm}|>
{\arraybackslash}m{3.8cm}|>
{\arraybackslash}m{4.8cm}|>
{\arraybackslash}m{4.8cm}|}
\rowcolor{verylightgray} \hline 

\multicolumn{1}{|c|}{\textbf{Feature}} & \multicolumn{1}{c|}{\textbf{5G}} & \multicolumn{1}{c|}{\textbf{6G}} & \multicolumn{1}{c|}{\textbf{Key differences in 6G}} & \multicolumn{1}{c|}{\textbf{Cybersecurity implications of 6G}} & \multicolumn{1}{c|}{\textbf{Application scenarios for 6G-ITS}} \\ [0.3ex]
\hline 
Data Transmission Speeds & 
Up to 20 Gbps & 
Up to 1 Tbps and beyond & 
6G enhances speeds by over 50x, which is crucial for massive ITS data needs. & 
Increased speeds require new protocols to ensure data integrity and timely encryption. & 
Remote Driving and Real-Time Telematics: Real-time vehicle data handling for autonomous control. \\
\hline
Latency & 
Around 1 ms & 
Sub-1 ms & 
Reduction in latency by up to 90\%, crucial for safety-critical operations in ITS. & 
Lower latency may expose systems to timing attacks unless security is adapted to operate within tighter time frames. & 
Autonomous Vehicle Coordination: Enables instant decision-making necessary for vehicle platooning. \\
\hline
Network Density & 
1 million devices/km² & 
10 million devices/km² & 
10x increase in device connectivity, supporting dense urban ITS deployments. & 
Higher device density intensifies the risk of DDoS attacks, necessitating more robust network defences. & 
Smart City Infrastructure: Comprehensive ITS deployment integrating traffic, safety, and emergency services. \\
\hline
Spectrum Efficiency & 
Utilizes up to 30 GHz & 
Expands to sub-terahertz bands (above 300 GHz) & 
Higher frequency bands increase bandwidth and connection density. & 
The expanded spectrum introduces a potential for more sophisticated eavesdropping and interference techniques. & 
AR and VR Services: Enhanced traffic management and navigation aids for drivers and public systems. \\
\hline
Reliability & 
99.999\% availability & 
99.99999\% availability & 
Enhanced network reliability through advanced technologies and AI integration. & 
Improved reliability reduces the risk of connection drops, which is crucial for maintaining continuous ITS operations. & 
Critical Emergency Response: Reliable communication for real-time coordination during safety incidents. \\
\hline
Security and Privacy & 
AES encryption standard & 
Quantum cryptography solutions & 
Implementation of quantum-resistant technologies ensuring superior security levels. & 
The introduction of quantum cryptography enhances security but requires updates to existing security frameworks to address new vulnerabilities. & 
Secure V2X Communication: More secure vehicle-to-everything communications for safer transportation systems. \\
\hline
Edge Computing & 
Latency-dependent processing & 
Edge AI with real-time processing capabilities & 
Improved processing capabilities at the edge are crucial for immediate data handling. & 
Enhanced edge computing capabilities necessitate advanced localized security measures to protect data at the edge. & 
Localized Data Processing: Real-time traffic data management to enhance flow and reduce congestion. \\
\hline
Device-to-Device Communication & 
Standard direct connectivity & 
Ultra-reliable direct device communications & 
Facilitates more robust and efficient device-to-device interactions. & 
Improved direct communications reduce reliance on central servers, shifting security focus to endpoint integrity. & 
Peer-to-Peer Networks for ITS: Direct vehicle communication enhances efficiency and reduces central dependency. \\
\hline
Interference Management & 
Managed with beamforming & 
Advanced AI-driven dynamic interference management & 
AI dynamically adapts to interference, enhancing communication quality. & 
Dynamic interference management may introduce vulnerabilities if AI systems are compromised. & 
Dynamic Spectrum Management: Active spectrum management to minimize interference in urban areas. \\
\hline
Network Slicing & 
Static slices tailored for specific services & 
Dynamic slicing with AI-driven adjustments & 
More adaptable network slices dynamically meet changing ITS demands.& 
Dynamic slicing increases the complexity of security management, requiring continuous monitoring and adaptation of security policies. & 
Dedicated Slices for Public Transit: High-priority network resources for transit systems ensure reliability and efficiency.\\
\hline
\end{tabular}%
}
\end{table*}
}

\subsection{6G-ITS Challenges}\label{sec:6gdiscussion}
One of the main challenges that were identified during the survey is the scalability and flexibility of the existing authentication systems and their adaptability to ITS.
As the number of entities interacting with the system increases, it becomes more difficult to manage and secure and ensure that the right level of access is granted to the right components. However, the major contention is the increased number of loosely connected devices 6G-ITS will support. To address this challenge, it is important to use secure and flexible authentication methods, such as multi-factor authentication, digital certificates, and PKI, that can scale to accommodate many entities without compromising security.

A major drawback for connecting several ITS components in a 6G environment is the authentication method currently being applied in VANET. 6G enabling technologies like IRS, VLC, Terahertz and intelligent integrated computing (cloud and edge) will create an additional connectivity spectrum for more cyber-physical systems. For example, VLC will enable features like vehicle headlights and traffic lights to communicate with vehicles and other ITS components within the vehicular network. Also, live traffic update reports to vehicles can be delivered through IRS-enabled communication to base stations and other vehicles in the vehicular network. Therefore, a centralized structure such as the Trusted Authority (TA) scheme, mostly used by current authentication protocols, would prove inefficient and ineffective in maintaining data integrity and privacy security in real-time in such a mobile, geographically distributed and heterogeneous network environment. Likewise, the authentication and key agreement mechanism (AKA) present in third-generation partnership projects (3GPP) in 4G and 5G are insufficient to manage several authentication requests of vehicles at a time. Attempts were made to address this problem. 

Ouaissa et al.~\cite{84} proposed an enhanced group authentication protocol for vehicular communications in 5G using the 5G-AKA and elliptic Curve Diffie-Hellman algorithm. However, the varying security levels of ITS components mean a symmetric cryptosystem utilized poses an inherent vulnerability of transmitting the same key needed to encrypt and decrypt. 

Li et al.~\cite{85} proposed a platoon handover authentication scheme in 5G-V2X. The proposed scheme leveraged the Access and Mobility Management function of the 5G core network and created two platoon handover authentication schemes within the Access and Mobility Management Function (AMF) called inter-AMF and intra-AMF. The platoons authenticate with a software network controller when a platoon enters the coverage area of a target gNB from a source gNB and achieves the handover threshold. The gNBs in the 5G-Radio Access Network (5G-RAN) carry out the platoon handover authentication. This scheme can prevent man-in-the-middle and replay attacks since messages carry timestamps. However, protecting the privacy of platoon members and streamlining the signalling process while the SDN controller monitors the platoon's position is one of the issues associated with the proposed approach because it may be using untrustworthy third-party applications. How platoons maintain seamless handover authentication between gNB and eNB or non-3GPP network access points is another issue. Authentication on other components of ITS, such as RSU, was also investigated.

Feng et al.~\cite{86} introduced a protocol known as P2BA, which emphasizes privacy and incorporates batch authentication against semi-trusted RSUs. When a registered vehicle communicates a traffic-related message and its concealed certificate to an RSU, the latter, using non-interactive zero-knowledge proof techniques, can independently ascertain the message's validity. Compared to anonymous authentication systems, P2BA demonstrated a remarkable reduction in both computational time and storage requirements. A salient feature of this protocol is its ability to ensure message integrity and nonrepudiation. In cases of disputes, only law enforcement can unveil the vehicle linked to the verification message, thus disclosing its real identity. Nevertheless, there might be efficiency challenges in heterogeneous networks. 

Recently, Kovalev et al.~\cite{87} investigated an authentication mechanism for VANETs reliant on RSU infrastructure. This mechanism authenticates message signatures in RSUs utilising elliptic curve cryptography. While it potentially reduces computational overhead and the amount of data exchanged, it carries risks such as a single point of failure and possible delays in communications between RSUs and core networks or among RSUs.  

More recently, Yang et al.~\cite{88} proposed a decentralized mutual authentication protocol with edge assistance. This protocol enables a swift, one-round, interaction-based authentication between vehicles and edge nodes. Several edge nodes can collaboratively validate each vehicle during the initial phase. These nodes subsequently provide an access token based on the vehicle's unique threshold signature, which can be used later for rapid handover authentication. This approach appears promising for highly mobile and dispersed settings, though there might be concerns regarding its computational efficiency.

To address these concerns, Tan et al.~\cite{89} proposed a solution that addresses the challenges of authenticating devices and vehicles in a distributed, infrastructure-less environment. The scheme is based on the idea that vehicles can act as mobile authentication servers and assist in authenticating other vehicles and devices in the network. The scheme begins with the registration of vehicles and devices, which includes the generation and distribution of digital certificates for the devices. During the authentication process, vehicles broadcast a challenge message and request a response from other devices in the network. The devices then use their digital certificates to respond to the challenge and prove their identity to the vehicle-assisted authentication server. The scheme also includes a trust management mechanism, which allows the authentication server to establish trust relationships with other vehicles and devices in the network based on their previous authentication history and reputation. This scheme is particularly useful in infrastructure-less environments such as rural or disaster-stricken areas, where the traditional fixed infrastructure is unavailable or damaged. The scheme enables the vehicles to act as a moving infrastructure, providing authentication and secure communication in a distributed environment.

Implementation of intelligent zero-trust technology in 6G-ITS can equally provide authentication solutions to components. Even after ITS components are authenticated and authorized, zero trust architecture continues to deliver network confidentiality and integrity under the presumption that no entity or component requesting connectivity or access to the network is considered safe and trusted~\cite{90}. Every access request is uniquely considered and approved in guidance with the security policy requirements for the trust evaluation. 

\begin{figure}[b]
    \centering
    \includegraphics[width=0.5\textwidth]{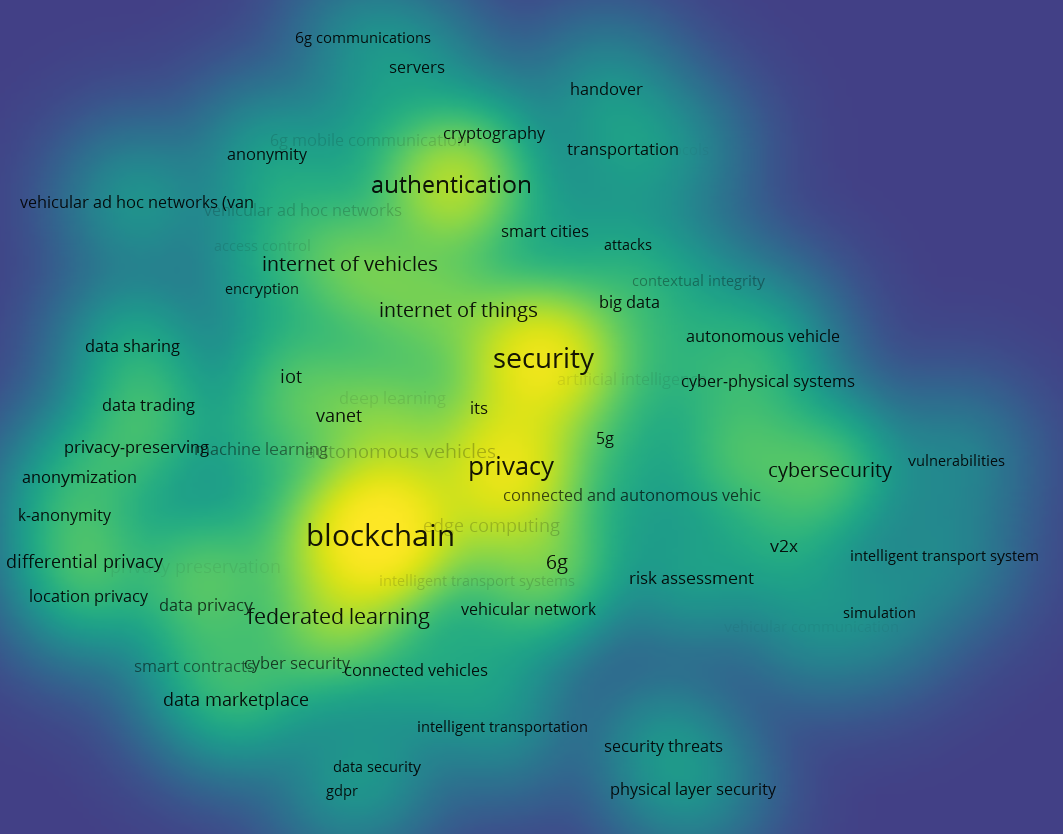}
    \caption{Surveyed paper keyword density.}
    \label{fig:density}
\end{figure}

Critical communication and tactile edge networks comprising heterogeneous and ubiquitous devices require next-generation networks like 6G~\cite{91}. Through a variety of new radio access technologies (RATs), including space air and ground integrated network (SAGIN), 6G can provide the required computational resources (through cloud computing) and seamless, dependable, and resilient connection~\cite{33}. Meng et al.~\cite{92} work on continuous authentication protocol without trust authority for zero-trust architecture that can be adapted to vehicular networks and other ITS. The paper employed blockchain to eliminate the trusted node, thereby decentralising the authentication, particularly for device-to-device continuous authentication. Additionally, Song et al.~\cite{93} proposed a new zero-trust-aided smart key authentication scheme on the IoV. The scheme proposed a smart key for users to unlock vehicles. The scheme proposed a continuous authentication system based on fingerprint, NFC, and facial data to authenticate the driver while driving. It should be noted that the scheme has not yet been implemented on a large scale, and this approach is still being researched. Before implementing in real-world scenarios, further studies should be conducted on security, scalability, and fault tolerance. Figure~\ref{fig:density} shows the authentication. Security, privacy, and blockchain are the keywords most discussed in the papers surveyed for this paper. 

\subsection{Implementation Barriers}
The transition from theoretical frameworks to the practical implementation of security, privacy, and trust mechanisms within 6G-ITS faces several implementation barriers. One primary challenge is the inherent complexity of integrating heterogeneous devices and technologies into cohesive, secure systems~\cite{10845830}. The coexistence of legacy systems with emerging technologies such as IRS, VLC, and terahertz communication complicates interoperability and standardization, leading to fragmented security protocols and inconsistent trust enforcement~\cite{9790832}. Furthermore, current authentication and key management schemes are insufficiently scalable or flexible to handle the anticipated density and dynamic interactions of 6G-ITS components. Traditional centralized approaches, such as the TA model, face limitations regarding latency, computational overhead, and vulnerability to single-point failures, necessitating a shift towards decentralized architectures~\cite{10819009}.

Decentralized approaches, such as blockchain-based identity management and distributed authentication protocols, address these traditional limitations, introducing new challenges that contribute to implementation barriers. These include increased communication overhead due to consensus mechanisms, synchronization difficulties in highly mobile environments, and scalability issues in real-time scenarios~\cite{10376064}. Moreover, the computational and storage demands of decentralized solutions can strain the resource-constrained edge devices commonly used in ITS. Importantly, many of these decentralized frameworks have only been validated in controlled simulation environments, and their performance can degrade or behave unpredictably under real-world conditions, where factors such as network instability, node failure, and physical interference are more difficult to control~\cite{9790832}.

Regulatory and governance issues present another critical barrier, particularly concerning data privacy compliance and cross-border interoperability. Differences in global regulations on data protection, such as GDPR in Europe, create complexities in the design of universally compliant privacy preservation mechanisms~\cite{10373408}. Furthermore, implementing sophisticated cryptographic techniques, including quantum-resistant algorithms, requires substantial computational resources and incurs significant deployment costs~\cite{71}. Real-world constraints, such as limited computational power and energy resources in edge devices, exacerbate these challenges, posing practical difficulties in achieving the desired security levels without degrading system performance.

Lastly, societal acceptance and ethics are subtle but influential obstacles. Public trust in autonomous transportation systems is highly dependent on transparent decision-making~\cite{10373556}. Ethical issues related to AI decisions, such as the trade-off between privacy and safety, and the accountability of automation, remain unresolved, potentially causing delays in the adoption of technology. Comprehensive research, international standards, and clear ethical guidelines are essential for the successful deployment and acceptance of secure 6G-enabled ITS.
\section{DISCUSSION}\label{sec:dis}
This section examines the findings of our analysis and maps them to the original research questions. Much of the current discourse around 6G-ITS remains conceptual. We critically explore how the frameworks for security, privacy, and trust have been validated in recent empirical studies. Drawing from both simulations and early-stage deployments, we assess the translation of theoretical models into practical implementations and identify instances where empirical evidence supports or contradicts design assumptions. This approach allows us to highlight both the maturity of certain technologies and the areas where real-world applicability is limited.
\subsection{Empirical Validation}
The theoretical foundations of the 6G-ITS security, privacy, and trust frameworks are extensively developed in the academic literature; however, empirical validation is relatively sparse. Most studies rely predominantly on high-fidelity simulations instead of deployment-grade field trials, resulting in a persistent gap between conceptual models and practical implementation~\cite{10,14,20,32,63,56,59,43,10373556}. This limitation is of particular importance for ITS, as real-world operational complexity, mobility patterns, and environmental variability can considerably affect the effectiveness of proposed mechanisms.

Several advanced testbeds provide partial empirical insight. The VIAVI GPU-Powered Real-Time 6G Testbed, developed in collaboration with the Institute for Wireless Internet of Things and the Open6G Cooperative Research Center at Northeastern University, integrates a city-scale digital twin to evaluate the 6G network capabilities~\cite{viavi2024_6g_testbed}. Similarly, the IEEE 5G/6G Innovation Testbed offers a cloud-based platform to assess emerging services in domains such as smart cities, industrial automation, and transportation~\cite{ieee2023testbed}. However, these environments primarily validate communication performance, such as throughput, latency, and reliability, rather than conducting end-to-end assessments of security, privacy, and trust mechanisms.

Empirical studies focused on security for ITS remain focused on 5G and VANET environments. For example, standalone core servers combined with software-defined radio (SDR) cards have been used to identify vulnerabilities through fuzzing operations, and the performance of the V2X protocol has been benchmarked in terms of latency and computational overhead~\cite{fujita2021terahertz}. However, there is little equivalent work for ITS enabled by 6G, particularly in the areas of quantum-resistant cryptography, zero-trust architectures, and AI-driven intrusion detection in vehicular contexts~\cite{10807044}. Validations of the existing trust models in vehicular ad hoc networks have been carried out using simulators such as MobiSim and NS-2, typically with limited mobility scenarios and without the integration of specific features of 6G, including submillisecond latency and ultra-dense connectivity~\cite{90,91,9,117,123}.

For privacy, empirical work is sparse. Machine learning–based channel estimation in simulated 6G conditions has achieved over 92\% accuracy in differentiating simultaneous users under moderate turbulence, and bidirectional LSTM models have reached 99.99\% accuracy in detecting DoS / DDoS attacks in 5G network slices~\cite{136}. Although these findings are promising for privacy-preserving federated learning and AI-driven intrusion detection, comprehensive evaluations of differential privacy, homomorphic encryption, and secure multiparty computation in vehicular networks remain confined to controlled simulations~\cite{160}. Trust frameworks for autonomous decision making also lack empirical implementation, with no large-scale evaluation of verifiable AI models in safety-critical ITS scenarios.

The current empirical landscape therefore reveals three critical gaps: (i) the absence of large-scale field trials that jointly validate security, privacy, and trust mechanisms under realistic 6G-ITS conditions, (ii) the lack of standardized benchmarks and metrics for consistent cross-scenario evaluation, and (iii) insufficient real-world testing of emerging technologies such as quantum-safe V2X authentication, verifiable AI, and longitudinal trust adaptation in adversarial multiagent environments. Furthermore, most validation efforts are siloed, focusing on either communication performance or isolated security features, without addressing their combined effects in integrated ITS deployments. These limitations highlight the urgent need for coordinated industry–academia efforts to establish empirical frameworks capable of bridging the theory–practice divide. As a result, several open problems remain, as summarized in Table~\ref{tab:open_problems}.

{
\renewcommand{\arraystretch}{1.4}
\begin{table*}[htbp]
\centering
\caption{Open problems in 6G-ITS security, privacy, and trust}
\label{tab:open_problems}
\resizebox{\linewidth}{!}{

\begin{tabular}{|>{\centering\arraybackslash}m{1.1cm}|>{\arraybackslash}m{5.67cm}|>
{\arraybackslash}m{11.2cm}|}
\hline 
 
\rowcolor{verylightgray}
\textbf{Category} & \multicolumn{1}{c|}{\textbf{Open problem}} & \multicolumn{1}{c|}{\textbf{Description}} \\ [0.3ex]

\hline

\multirow{2}{*}{\vspace{-0.4cm} Security}  & Zero-trust enforcement for O-RAN-based ITS & Applying zero-trust models at the RIC and edge nodes is still an open challenge due to latency constraints and difficulty in synchronising dynamic policies across distributed components. \\ \cline{2-3}

 & Quantum-safe V2X authentication & Post-quantum V2X protocols remain underdeveloped for real-time, mobile vehicular environments, with limited field validation of algorithm performance under sub-millisecond latency requirements. \\ \hline

\multirow{2}{*}{\vspace{-0.4cm} Privacy} & Cross-domain data minimisation & Balancing data utility and privacy across ITS, telecom, and cloud domains—particularly under federated learning constraints—remains unresolved in real-world deployments. \\ \cline{2-3}

 & Differential privacy for real-time sensor data & Most differential privacy mechanisms degrade performance or fail to meet latency requirements for safety-critical 6G-ITS applications. \\ \hline

\multirow{2}{*}{\vspace{-0.4cm} Trust} & Verifiable AI in safety-critical decisions & ITS lacks verifiable and interpretable ML models that maintain robustness under adversarial conditions while providing real-time decision assurance in safety-critical contexts. \\ \cline{2-3}

& Dynamic trust computation in multi-agent systems & Computing and adapting trust scores for vehicles in dynamic, adversarial, and high-mobility settings remains underexplored, particularly for large-scale, real-world ITS deployments. \\ \hline
\end{tabular}
}
\end{table*}
}

\subsection{RQ1:In the context of 6G-ITS, how can multi-layered security strategies be designed to protect against sophisticated cyberattacks targeting communication networks, devices, and data analytics platforms?} 
Addressing the challenge of crafting multi-layered security strategies for 6G-ITS, our investigation highlighted the necessity for robust frameworks capable of safeguarding these systems against an evolving landscape of cyber threats. The advent of 6G technology brings forth enhanced data rates, connectivity, and reduced latency, marking significant advancements in transportation efficiency and safety. However, these benefits also introduce complex security vulnerabilities, from increased attack surfaces due to the sheer volume of connected devices to sophisticated cyberattacks that disrupt transportation systems' integrity and availability. Our discussion on multi-layered security strategies underscored the integration of cutting-edge encryption techniques, secure communication protocols, and stringent access controls, all tailored to shield ITS against potential breaches. Notably, the prospect of quantum computing necessitates a forward-looking approach to ITS security, prompting the inclusion of quantum-resistant algorithms in the strategic defence matrix. By synthesizing theoretical models with real-world case studies, the paper presents actionable solutions that fortify the security posture of ITS, ensuring a resilient infrastructure capable of withstanding future cyber threats.

\subsection{RQ2: What mechanisms can be developed within 6G-ITS to ensure user privacy in the face of extensive data collection required for advanced sensing, automation, and communication systems?}
The discourse on privacy-preserving mechanisms within 6G-ITS illuminated the critical balance between leveraging extensive data for system optimization and safeguarding individual privacy. With the advent of 6G, ITS is poised to benefit from unprecedented data collection and processing levels, enhancing traffic management and vehicle-to-everything communications. However, this capability raises significant privacy concerns, necessitating the adoption of technologies like federated learning and differential privacy. These methodologies enable the decentralized analysis of data, minimizing the exposure of sensitive information while maintaining system efficacy. The paper further explores the regulatory and ethical frameworks essential to embedding privacy considerations into ITS design and operation. It advocates for a privacy-centric approach that aligns with global standards and public expectations.

\subsection{RQ3: How can trust in 6G-ITS be quantified and managed, particularly in systems involving decision-making and autonomous operations?}
Lastly, our inquiry into quantifying and managing trust in AI/ML-driven ITS systems revealed the complexities involved in ensuring the public perceives these technologies as reliable and safe. The deployment of AI/ML in autonomous vehicle operations and decision-making processes introduces a layer of opacity that can hinder trust. This research underscores the importance of developing transparent and verifiable methods to evaluate the reliability of AI/ML algorithms and their output. Establishing trust in ITS extends beyond technical solutions to encompass ethical considerations, stakeholder participation, and adherence to the principles of responsible AI usage. The paper advocates for a comprehensive trust management framework, incorporating standardized metrics for trust quantification alongside adaptive strategies that evolve with technological advancements and societal values.
\section{CONCLUSION}\label{sec:conclusion}
This comprehensive survey reveals that while 6G-ITS promises transformative improvements in connectivity, automation, and safety, significant challenges in security, privacy, and trust must be resolved before widespread deployment. Our analysis identifies critical gaps between theoretical frameworks and practical implementation, particularly in empirical validation, quantum-safe integration, and real-world scalability testing. The evolution from 5G to 6G fundamentally alters the security paradigm, requiring distributed defense mechanisms, quantum-resistant cryptography, and AI-driven threat detection that current approaches cannot adequately address.

To bridge these gaps, coordinated action between stakeholders is essential. Regulators must establish standardized security-by-design mandates with clear post-quantum migration timelines and cross-border data governance frameworks. The industry must embed zero-trust architectures, hardware security foundations, and privacy-preserving analytics throughout its development lifecycles. Network operators require end-to-end protection that combines slice isolation, confidential computing, and robust incident response capabilities. Research institutions must advance measurable trust metrics, formal verification methods, and empirical validation through joint pilot programs that bridge the gaps between lab and field implementation. Success in 6G-ITS deployment depends on establishing technically rigorous, privacy-preserving security baselines that are validated through comprehensive real-world testing and supported by adaptive governance frameworks that evolve in response to emerging threats.
\balance
\bibliographystyle{ieeetr}
\bibliography{interplay}
\end{document}